\documentclass[article,12pt]{article}

\usepackage{amscd,amsmath,amssymb,amsfonts,latexsym,mathrsfs,amsthm,mathtools}
\usepackage{graphicx} 
\usepackage{bigints}
\usepackage{setspace} 
\usepackage{graphics,epsfig}
\usepackage{sectsty}
\usepackage[english]{babel}
\usepackage{bm}
\usepackage{multirow}
\usepackage{subfigure}
\usepackage[usenames]{color}
\usepackage{rotating}
\usepackage{url}
\usepackage{bbm}
\usepackage{float}
\usepackage{tikz}
\usepackage{pgf}
\usepackage{pdflscape}
\usepackage{xcolor}
\usepackage{colortbl}
\usepackage{bigints}

\newtheorem{rem}{{\bf Remark}}

\usepackage{flushend}
\usepackage{verbatim}
\usepackage{tabularx}
\usepackage{arydshln}
\usepackage{hyperref}

\newcommand{\E}{\mathbb{E}}

\newcommand{\x}{\bm{\theta}}

\newcommand{\y}{{\bf y}}

\newcommand{\post}{{\bar \pi}}

\definecolor{MYCOLOR0}{rgb}{0.92,0.92,0.92}
\definecolor{MYCOLOR}{rgb}{1,1,0}
\definecolor{MYCOLOR2}{rgb}{0.5,1,0.5}
\definecolor{MYCOLOR3}{rgb}{0.88,1,1}

\usepackage{imakeidx}
\makeindex

\UseRawInputEncoding

\title{Nested Sampling: A Critical and Comprehensive Theoretical Guide} 

\author{L. Martino$^\star$$^\dagger$\footnote{The authors have no conflicts of interest. Any data and code used in this work will be made publicly available upon publication. }, F. Llorente$^\top$, \\
{\small$^\star$  Universit{\'a} di Catania, Catania (Italy). }\\
{\small $^\top$ Stony Brook University, New York (USA).} \\
{\small $^\dagger$ corresponding author: luca.martino@unict.it.}
}

\date{2026}
 
\begin{document}

\maketitle

\thispagestyle{empty}

\begin{abstract}
The nested sampling (NS) technique has gained widespread attention, particularly in cosmology and astronomy, due to its ability to efficiently explore high-likelihood regions - a feature akin to an implicit likelihood optimization that underlies its success. While the full theoretical derivation of NS is complex and involves several approximations, the central challenge lies in sampling from the likelihood-constrained priors, which is crucial for its performance. This work provides a comprehensive and detailed exposition of NS derivation, clarifying both its theoretical foundations and practical challenges. 
 We provide a thorough description of the NS procedure, emphasizing both its strengths and potential limitations. { The connections between NS, importance sampling (IS), and vertical Monte Carlo are also discussed in detail.}
 In doing so, this work seeks to deepen understanding of the method and to foster the development of future enhancements, novel variants, and more efficient implementations across a wide range of scientific applications. Thus, the main contribution of this work is twofold: it serves both as a tutorial for newcomers to the field and as a critical review for experienced practitioners.
\end{abstract}

\section{Introduction}
\label{sec-intro}
Nested Sampling (NS) \cite{skilling2004nested,skilling2006nested} is a stochastic quadrature method designed to approximate high-dimensional and potentially complex integrals \cite{Liu04b,martino2018independent}. The main framework of application is Bayesian inference \cite{gelman2013bayesian,Liu04b,Robert04}. Together with the families of importance sampling (IS) and Markov chain Monte Carlo (MCMC) schemes, NS has become a central tool in modern computational statistics and its scientific applications \cite{Buchner23,chopin2010properties}. {Excellent, comprehensive and up-to-date reviews of the methodology and its developments can be found in \cite{Ashton2022,Buchner23,Keskin2023,Latz2024NSUncertainty}.}
IS techniques and their sequential  versions (also known as particle filters) are widely used in engineering applications, while MCMC algorithms have long dominated the field of statistics. NS, on the other hand, has achieved particular prominence in the physical sciences, especially in cosmology and astronomy \cite{Ashton2022,Buchner23,Mukherjee_2006} {and, more recently, also in engineering and statistical signal processing \cite{Keskin2023}.}
Like IS schemes - and in contrast to standard MCMC algorithms -  NS is able to simultaneously provide estimates of model parameters and an approximation of the marginal likelihood \cite{knuth2015bayesian,llorenteREV1,zhao2017integrated}. 
Indeed, originally NS was introduced precisely for the purpose of computing the marginal likelihood (a.k.a., the Bayesian evidence) \cite{skilling2004nested}. Over the past two decades, substantial progress has been made in understanding the theoretical foundations of the algorithm and in elucidating its relationships with other computational methodologies  \cite{chopin2010properties}. During this time, numerous efficient implementations, methodological refinements, and diagnostic tools have been proposed to enhance its reliability and performance \cite{feroz2008multinest,Latz2024NSUncertainty,Keskin2023}. As a result, the range of applications of NS has expanded well beyond its original applications in cosmology, finding use across a broad spectrum of scientific disciplines \cite{Keskin2023}.
\newline
This work aims to provide a comprehensive understanding of the NS procedure for all the possible readers, highlighting both its strengths and potential critical issues. Indeed, although the success of NS is undeniable and widespread, we find it somewhat remarkable (and surprising) when one carefully examines the full derivation of the method. 
 In particular, several authors have pointed out that the derivation of the method relies on a number of approximations, which can affect the statistical properties of the resulting estimators \cite{chopin2007,chopin2007b,chopin2010properties,Higson2018}. Motivated by these observations, this work not only presents the methodology in a pedagogical manner but also examines its underlying assumptions and discusses the principal criticisms and theoretical challenges that have emerged in the statistical literature.
\newline
 The overall NS procedure contains an implicit optimization of the likelihood function that is{, in our opinion,} the key of the NS success. {Another strength of NS is that its theoretical foundations provide clear diagnostic criteria, whose validity, however, relies on the satisfaction of the main assumptions underlying the theory. However, all the NS theory relies on the idea of sampling from} {\it likelihood-constrained prior densities}, denoted as $g(\x|\lambda)$. In our view, sampling from $g(\x|\lambda)$ can be even more challenging than sampling directly from the posterior. To illustrate this point, note that determining the support of these truncated priors would, in principle, require inverting the likelihood function, that is a task that is generally infeasible or computationally prohibitive. 
 This issue has been emphasized by prominent authors in computational statistics \cite{chopin2007,chopin2007b}, who note that {\it ``in high-dimensional spaces, simulating from the prior until the constraint is satisfied is unrealistic''.} The same authors also discuss the potential poor performance when using vague priors, and the complete impracticality of the method in the case of improper priors \cite{chopin2007,chopin2007b}. The problem of sampling within domains constrained by likelihood values is, in principle, the same theoretical challenge encountered in {\it slice sampling} \cite{NealSlice}. In both cases, NS and slice sampling, the development of robust and efficient computational implementations, often incorporating sophisticated internal Monte Carlo (MC) techniques to draw from $g(\x | \lambda)$, appears to have effectively ``solved'' the problem, at least from a practical point of view.  Another insightful theoretical analysis of NS is presented in \cite{Higson2018}. The authors identify and investigate two principal sources of error in NS, both arising from the approximations underlying the derivation of the method. These issues highlight some of the inherent limitations of NS and are closely related to several aspects that we seek to emphasize throughout this work. A complete advanced theoretical treatments of nested sampling from a statistical standpoint are provided in \cite{chopin2010properties,Higson2018,Keeton11}.
\newline
Thus, the main contribution of this work is twofold. First, it is intended as a tutorial for readers who are new to nested sampling, providing  a self-contained introduction with all details of its derivation. Second, it offers a critical review aimed at experienced practitioners, highlighting both the strengths and the limitations of the approach. {Moreover, we provide a detailed analysis of the connections among nested sampling, importance sampling (IS), and vertical Monte Carlo \cite{polson2014vertical}. We also discuss possible ways of combining these approaches to exploit their complementary strengths. To the best of our knowledge, some aspects of this discussion can constitute novel contributions to the literature.}
\newline
The remainder of the paper is organized as follows. Section~\ref{NotSect} introduces the necessary background material and establishes the main notation. The theoretical framework required for a rigorous description of the NS procedure is developed in Sections~\ref{OneIntRep} and~\ref{Add_teo_NS}. A comprehensive and detailed presentation of the Nested Sampling (NS) algorithm is provided in Section~\ref{NestedSampling}. Its connections with importance sampling (IS), as well as more advanced NS variants, are examined in Sections~\ref{SuperSect} and~\ref{SolSect}. Finally, concluding remarks are presented in Section~\ref{SectConc}.


\section{Background and main notation}\label{NotSect}
In many applications, the goal is to make inference about a variable of interest, $\boldsymbol{\theta}= \left[\theta_1, \theta_2, \ldots, \theta_{D_\theta}\right] \in \Theta \subseteq \mathbb{R}^{D_\theta}$, where $\theta_d \in \mathbb{R}$ for all $d=1, \ldots, D_\theta$, given a set of observed measurements, $\mathbf{y}=\left[y_1, \ldots, y_{D_y}\right] \in \mathbb{R}^{D_y}$. { We assume that $\Theta\subseteq \mathbb{R}^{D_\theta}$ is a connected parameter space.} In a standard Bayesian framework, we assume a known observation model that induces a likelihood function $\ell(\mathbf{y} | \boldsymbol{\theta})$.  Assuming a prior probability density function (pdf) $g(\boldsymbol{\theta})$ over the parameter vector to infer,  all the statistical information is summarized by the posterior density, i.e.,
\begin{align}
\post(\boldsymbol{\theta})=p({\bm \theta}|{\bf y})=\frac{\ell(\mathbf{y} | \boldsymbol{\theta}) g(\boldsymbol{\theta} )}{p(\mathbf{y})} {\propto \pi(\x)=\ell(\mathbf{y} | \boldsymbol{\theta}) g(\boldsymbol{\theta}),}
\end{align}
where 
\begin{align}
Z_{\texttt{ev}}=p(\mathbf{y})&=\int_{\Theta} \ell(\mathbf{y} | \boldsymbol{\theta}) g(\boldsymbol{\theta} ) d \boldsymbol{\theta}=\int_{\Theta} \pi({\bm \theta}) d \boldsymbol{\theta},
\end{align}
is the so-called marginal likelihood, a.k.a., Bayesian {\it evidence}. This quantity is important for model-selection purposes, as we show below. However, usually $Z_{\texttt{ev}}=p(\mathbf{y})$ is unknown and difficult to approximate.  We can  evaluate the integrand, i.e., unnormalized target (posterior) function, $\pi(\boldsymbol{\theta})$. Clearly, note that $\post(\boldsymbol{\theta}) = \frac{1}{Z_{\texttt{ev}}} \pi(\boldsymbol{\theta})$. Several methods have been proposed to compute the marginal likelihood $Z_{\texttt{ev}}$ , and virtually all of them rely on importance sampling (IS) identities \cite{chib1995marginal,chen2005computing,friel2008marginal,llorenteREV1}. In contrast, the nested sampling (NS) algorithm is primarily based on a different class of representations, often referred to as vertical likelihood representations \cite{polson2014vertical}. The connections and similarities between NS and related IS schemes are discussed in Section~\ref{SuperSect}.


\section{One-dimensional representations of the marginal likelihood}\label{OneIntRep}

In this section, we present an alternative approach based on the Lebesgue representation of the integral defining the marginal likelihood $Z_{\texttt{ev}}$ . We begin by deriving two equivalent one-dimensional integral formulations of $Z_{\texttt{ev}}$ , and then discuss how these representations can be exploited through the application of one-dimensional quadrature methods. The practical implementation of such quadrature rules, however, is not straightforward. An appropriate and carefully design of the nested sampling (NS) scheme will be therefore required to make their use effective.

\subsection{First one-dimensional representation {and main assumptions}}\label{Sect_IntegralsRep}
The $D_\theta$-dimensional integral $Z_{\texttt{ev}}=\int_\Theta\ell(\y|\x)g(\x)d\x$ can be turned into a one-dimensional integral using an extended space representation.  Namely, we can write 
\begin{align}
Z_{\texttt{ev}} &= \int_\Theta\ell(\y|\x)g(\x)d\x, \label{FirstOne} \\
&= \int_{\Theta}g(\x)d\x \int_0^{\ell(\y|\x)}d\lambda, \quad \text{(extended space representation)} \\
&= \int_{\Theta}g(\x)d\x \int_0^\infty \mathbb{I}\{0<\lambda<\ell(\y|\x)\}d\lambda,  
\end{align}
where $\mathbb{I}\{0<\lambda<\ell(\y|\x)\}$ is an indicator function which is $1$ if $\lambda \in [0,\ell(\y|\x)]$ and $0$ otherwise. Switching the integration order, we obtain 
\begin{align}\label{aquiHlam}
\boxed{
\begin{aligned}
Z_{\texttt{ev}} &= \int_0^\infty d\lambda\int_{\Theta}g(\x) \mathbb{I}\{0<\lambda<\ell(\y|\x)\}d\x \\
&= \int_0^\infty d\lambda\int_{\ell(\y|\x)>\lambda}g(\x)d\x \\
&= \int_0^\infty Z(\lambda)d\lambda, \\ 
&= \int_0^{{\sup \ell(\y|\x)}} Z(\lambda)d\lambda, 
\end{aligned}}
\end{align}
where we have set
\begin{align}\label{aquiHlam2}
\fbox{$ \displaystyle Z(\lambda) = \int_{\ell(\y|\x)>\lambda}g(\x)d\x =\mathbb{P}\left(\ell(\y|\x)>\lambda\right),\quad \mbox{ where } \quad \x \sim g(\x).$}
\end{align}

{\rem $Z(\lambda)$ represents a normalized area, i.e., $Z(\lambda) \in[0,1]$ (as shown in Figure \ref{figNested}).  Thus, we have $Z(0)=1$ and $Z(\lambda')=0$ for all  $\lambda' \geq {\sup\limits_{\x \in \Theta} \ell(\y|\x)}$,  and is also a non-increasing function, namely, is a {\it survival function}.}
\newline
\newline
{{\bf Assumptions.}  For simplicity of the exposition, we assume throughout that:
\begin{itemize}
\item The likelihood function $\ell(\y |\x)$ is continuous and continuously differentiable with respect to $\x$ on $\Theta$, where $\Theta$ is assumed to be a connected subset of $\mathbb{R}^{D_\theta}$.
\item  The likelihood function is always strictly positive, $\ell(\y |\x)>0$ for all $ \x \in \Theta$.
\item  The likelihood function $\ell(\y |\x)$ is not constant on $\Theta$. Thus, we avoid ``plateaus'', which would complicate the subsequent derivations. For further details, see \cite{Fowlie21plateaus}.
\item The prior density $g(\x)$ is strictly positive everywhere on $\Theta$, i.e., $g(\x)>0$ for all $\x \in\Theta$.
\item If $\x\sim g(\x)$, then the induced random variable $ L=\ell(\yÊ| \x)$ has a continuous distribution.
\item In Eq. \eqref{aquiHlam} and throughout the remainder of the work, we additionally assume that $\ell(\y| \x)$ is bounded on $\Theta$. Hence, the upper limit of integration can be taken to be
$\sup_{ \x\in\Theta}\ell(\y|\x)$, which is finite.
\end{itemize}
Some of these conditions can be easily relaxed by formulating them as ``almost everywhere'' statements, meaning that they could not hold on subsets of $\Theta$ of measure zero.} Below, we define several variables and sampling procedures required for the proper understanding of the nested sampling algorithm.

\subsection{Second one-dimensional representation}
Now let consider a specific {\it area} value $a=Z(\lambda)$ {(see Figure \ref{figNested} for two graphical examples). The generalized inverse function is then:} 
\begin{align}
\fbox{$ \displaystyle \Lambda(a) = Z^{-1}(a) { =\inf \{\lambda \geq 0: \  Z(\lambda)\leq a  \} \mbox{ with } a\in [0,1],} $}
\end{align} 
and is also non-increasing. {Under the above definition and assumptions, note that $Z(\lambda)>a$ if and only if $\lambda < \Lambda(a)$, i.e., $Z(\lambda)$. Moreover, we have the two extreme values:
$$
\Lambda(0)=\sup_{\x \in\Theta}\ell(\y| \x),
\qquad
\Lambda(1)=0.
$$
 i.e., $\Lambda(a): [0,1] \rightarrow \left[0, \sup\ell(\y | \x)\right]$. Moreover, under the above assumptions, $Z(\lambda)$ has neither jumps nor flat regions over the interior of its effective range,
 $\lambda\in\left[0,  \sup\ell(\y | \x)\right)$. }
  Then, we can write 
\begin{align}\label{SecondVertical}
\boxed{
\begin{aligned}
Z_{\texttt{ev}} &= \int_0^\infty Z(\lambda)d\lambda \\
&= \int_0^\infty d\lambda \int_0^1 \mathbb{I}\{a<Z(\lambda)\}da \qquad \text{(again the extended space ``trick'')}  \\
&= \int_0^1da \int_0^\infty\mathbb{I}\{a<Z(\lambda)\} d\lambda  \qquad \text{(switching  the integration order)}   \\
&= \int_0^1da \int_0^\infty \mathbb{I}\{\lambda<\Lambda(a)\}d\lambda \qquad \text{(using } Z(\lambda)>a \iff \lambda<\Lambda(a) \text{)}   \\
&=\int_0^1\Lambda(a)da. 
\end{aligned}}
\end{align}

{\rem Hence, we have obtained two of the marginal likelihood $Z_{\texttt{ev}}$  in Eqs. \eqref{aquiHlam}-\eqref{SecondVertical}, that is generally defined by an highly-multidimensional integral given in Eq. \eqref{FirstOne}.}
\newline
\newline
Note that functions $Z(\lambda)$ and $\Lambda(a)$ take into account both the prior density and likelihood  function. See Figure \ref{figNested0} for a graphical representation of these two one-dimensional functions.

\subsection{Discussion on possible quadratures}
\begin{figure}[!h]
	\centering
\includegraphics[width=15cm]{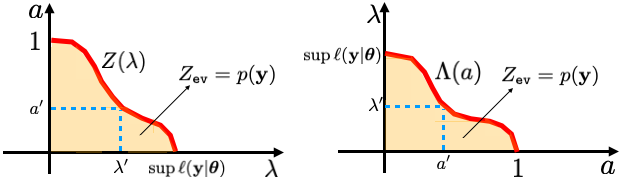}
	\caption{Graphical examples of the two one-dimensional functions $Z(\lambda)$ and $\Lambda(a)$. The area below the curve is in both cases the marginal likelihood $Z_{\texttt{ev}}$ . The marginal likelihood is generally expressed by a multi-dimensional integral, i.e., $Z_{\texttt{ev}}=\int_{\Theta} \ell(\mathbf{y} | \boldsymbol{\theta}) g(\boldsymbol{\theta} ) d \boldsymbol{\theta}$.
	}
	\label{figNested0}
\end{figure}

We have thus obtained two alternative one-dimensional integral representations of the Bayesian evidence $Z_{\texttt{ev}}$, which was originally defined as a multidimensional integral over the parameter space $\Theta$, i.e.,
\begin{align}\label{ZasOneDimIntegral}
\fbox{$\displaystyle Z_{\texttt{ev}} =\int_0^{{\sup \ell(\y|\x)}} Z(\lambda)d\lambda =\int_0^1 \Lambda(a)da.$}
\end{align}
The two integrand functions $a=Z(\lambda)$ and $\lambda=\Lambda(a)$ are graphically represented in Figure \ref{figNested0}.
{Hence, the Bayesian evidence $Z_{\texttt{ev}}$, originally defined as a multidimensional integral over $\Theta$, admits an equivalent representation as a one-dimensional integral over a bounded domain.} This reformulation can substantially simplify its numerical evaluation, and  we could think of applying simple quadrature: choose a grid of points in $[0,\sup\ell(\y|\x)]$ ($\lambda_i> \lambda_{i-1}$) or in $[0,1]$ ($a_i> a_{i-1}$), evaluate $Z(\lambda)$ or $\Lambda(a)$ and use the quadrature formulas of the form:
\begin{align}
	\widehat{Z}_{\texttt{ev}} &=  \sum_{i=1}^I (\lambda_i - \lambda_{i-1})Z(\lambda_i), \enskip \text{or} \\
	\widehat{Z}_{\texttt{ev}} &=  \sum_{i=1}^I (a_i - a_{i-1})\Lambda(a_i).
\end{align}
However, these simple approaches are difficult to be applied since (i) the functions $Z(\lambda)$ and $\Lambda(a)$ are intractable in most cases and (ii) they change much more rapidly over their domains than does $\pi(\x)=\ell(\y|\x)g(\x)$. {The behavior of the integrands at both boundaries (but especially near the maximum value of the likelihood function) contributes significantly to the value of the integral.} Hence the quadrature approximation can have very bad performance, unless the grid of points is chosen with extreme care. {For instance, $Z(\lambda)$ can be so sharply peaked that standard quadrature methods struggle to accurately evaluate the integral, even though it is only one-dimensional. Nested sampling (NS) can be viewed as a stochastic method that ``auto-magically'' constructs an effective grid for this quadrature.}
In the remainder of this work, we describe in detail the key concepts, additional variables, procedures, and results required for a proper introduction to the NS method. {Table \ref{SummaryNotation} summarizes the main notation of the work.}

\begin{table}[!h]
{
	\caption{Main notation of the work.}
	\label{SummaryNotation}
	\vspace{-0.5cm}
	\begin{center}
		\begin{tabular}{ll}  
		   	\hline 	
			$\ell(\y|\x)$  & likelihood function \\
			$g(\x)$  & prior density \\
		     $\pi(\x)=\ell(\y|\x)g(\x)$ & unnormalized posterior function \\
		                   $\post(\x)$ & posterior density, $\post(\x)=\frac{\pi(\x)}{Z_\texttt{ev}}$ \\
		  $Z_\texttt{ev}$ &  Bayesian evidence - marginal likelihood \\  
		  \hline 
	 $a$        & Normalized area within $[0,1]$   \\    
$\lambda$  & likelihood value (ordinate) and threshold value in $g(\x|\lambda)$  \\ 	                 
$a=Z(\lambda)$ & Constrained prior mass,  $\int_{\ell(\y|\x)>\lambda}g(\x)d\x =\mathbb{P}\left(\ell(\y|\x)>\lambda\right)$, $\x \sim g(\x)$    \\
$\lambda=\Lambda(a)$  &  inverse of $Z(\lambda)$, i.e., $Z^{-1}(a)$  \\
$L$ & random variable $L=\ell(\y|\x)$ with $\x \sim g(\x)$ \\
$F(\lambda)$  &  cumulative function of $L$, i.e.,  $F(\lambda)=1-Z(\lambda)=\mathbb{P}\left( \ell(\y|\x) \leq  \lambda\right)$,  $\x \sim g(\x)$   \\
$\widetilde{a}$ & ratio of normalized areas, e.g., $\widetilde{a}_{\texttt{max}}^{(i)}=\frac{a_{\texttt{max}}^{(i)}}{a_{\texttt{max}}^{(i-1)}}$, with $i$ is an iteration index.  \\
$g(\x|\lambda)$ & constrained prior pdf, defined as $\frac{1}{Z(\lambda)}\mathbb{I}\{\ell(\y|\x)>\lambda\} g(\x)$ \\
  $\mathcal{P}_i=\{\x_n\}_{n=1}^N$ & population of live points, such that $\x_n \sim g(\x|\lambda_{\texttt{min}}^{(i)})$ with $\x_n\in \mathcal{P}_i$   \\
   $\x_{\texttt{rem}}^{(i)}$ & removed point, defined as $\arg\min\limits_{\x\in \mathcal{P}_{i-1}}\ell(\y|\x)$, then we set $\mathcal{P}_i=\mathcal{P}_{i-1} \backslash \{\x_{\texttt{rem}}^{(i)}\}$. \\
    $\gamma_i$ & NS quadrature weight $\gamma_i=a_{\texttt{max}}^{(i-1)} -a_{\texttt{max}}^{(i)}\in (0,1)$ and  $\gamma_i\approx e^{-\frac{i-1}{N}}-e^{-\frac{i}{N}}$\\
			\hline
		\end{tabular}
	\end{center}
}
\end{table}

%
%
%
%

\section{Additional theoretical foundations of NS}\label{Add_teo_NS}

\subsection{On the survival function $ Z(\lambda)$ and related sampling procedures}\label{ZandSampling}

The function above $Z(\lambda): \mathbb{R}^{+}\rightarrow [0,1]$ is the mass of the prior restricted to the set $\{\x: \ell(\y|\x)>\lambda\}$.  Note also that
\begin{align}
\fbox{$\displaystyle Z(\lambda)=\mathbb{P}\left(\ell(\y|\x)>\lambda\right),\quad \mbox{ where } \quad \x \sim g(\x).$}
\end{align}
Moreover, we have that  $Z(\lambda) \in [0,1]$ with $Z(0)=1$ and $Z(\lambda')=0$ for all  $\lambda' \geq {\sup \ell(\y|\x)}$,  and it is also a non-increasing function. Therefore, $Z(\lambda)$  is a {\it survival function}, while  
\begin{eqnarray}
\fbox{$\displaystyle F(\lambda)=1-Z(\lambda)=\mathbb{P}\left( \ell(\y|\x) {\leq} \lambda\right)=\mathbb{P}\left(L{\leq} \lambda\right), \quad \mbox{ with } \quad \x \sim g(\x),$ } \label{EqF}
\end{eqnarray} 
is the cumulative function of a random variable $L=\ell(\y|\x)$ where $\x \sim g(\x)$ \cite{martino2018independent,Robert04} { (recall we assume the continuity of the distribution of the random variable $L$)}.  Note that $F(0)=0$ and $F(\lambda')=1$ for all  $\lambda' \geq {\sup \ell(\y|\x)}$.
{ Hence, $F(\lambda)$  is a  cumulative function and  $Z(\lambda)=1-F(\lambda)$ is a survival function, both representing normalized areas, i.e., probabilities.}

  \subsection{Sampling {from the cumulative functions $F(\lambda)$  and $ Z(\lambda)$}}\label{SectquiZ}
   The following procedure generates samples $\lambda_n$ from $F(\lambda)$ in Eq. \eqref{EqF}: 
\begin{enumerate}
\item Draw $\x_n\sim g(\x)$, for $n=1,...,N$.   
\item Set $\lambda_n=\ell(\y|\x_n)$,  for all $n=1,...,N$.
\end{enumerate}
{ Recalling the assumption of continuity of the distribution  of random variable $L=\ell(\y|\x)$ with $\x\sim g(\x)$,} and recalling the inversion method \cite[Chapter 2]{martino2018independent}, { we can write that the corresponding value $b_n=F(\lambda_n)$  is uniformly distributed in $[0,1]$ (since $F$ is the cumulative function of the random variable $L$, and $\lambda_n$ is a realization of this random variable), i.e.,}
\begin{eqnarray}
b_n=F(\lambda_n)\sim \mathcal{U}([0,1]).
\end{eqnarray}
 Recall that, if $U\sim \mathcal{U}([0,1])$, 
 $V=1-U$ is also uniformly distributed $\mathcal{U}([0,1])$, 
 then we also have
\begin{eqnarray} \label{AquiAn}
a_n=Z(\lambda_n)=1-F(\lambda_n)\sim \mathcal{U}([0,1]).
\end{eqnarray}
 In summary, finally we have that
\begin{equation}\label{AquiAn2}
\boxed{
\begin{aligned} 
 \mbox{ if }  & \quad \x_n \sim g(\x),   \\ 
 \mbox{ and } &\quad  \lambda_n=\ell(\y|\x_n) \sim F(\lambda){=1-Z(\lambda)},  \\
  \quad 
  \mbox{ then }  &\quad  a_n=Z(\lambda_n) \sim \mathcal{U}([0,1]). 
\end{aligned}
}
\end{equation}
{ All these expressions rely on the continuity of the distribution of the random variable $L$. Moreover, note that 
we can draw samples from $F(\lambda)$, but we cannot evaluate $F(\lambda)$ (or $Z(\lambda)$) because its analytical form is unknown. }
\subsection{The likelihood-truncated prior density  $g(\cdot|\lambda)$}\label{ZandSampling2} 
\begin{figure}[!h]
	\centering
\subfigure[\label{figNesteda}]{\includegraphics[width=7cm]{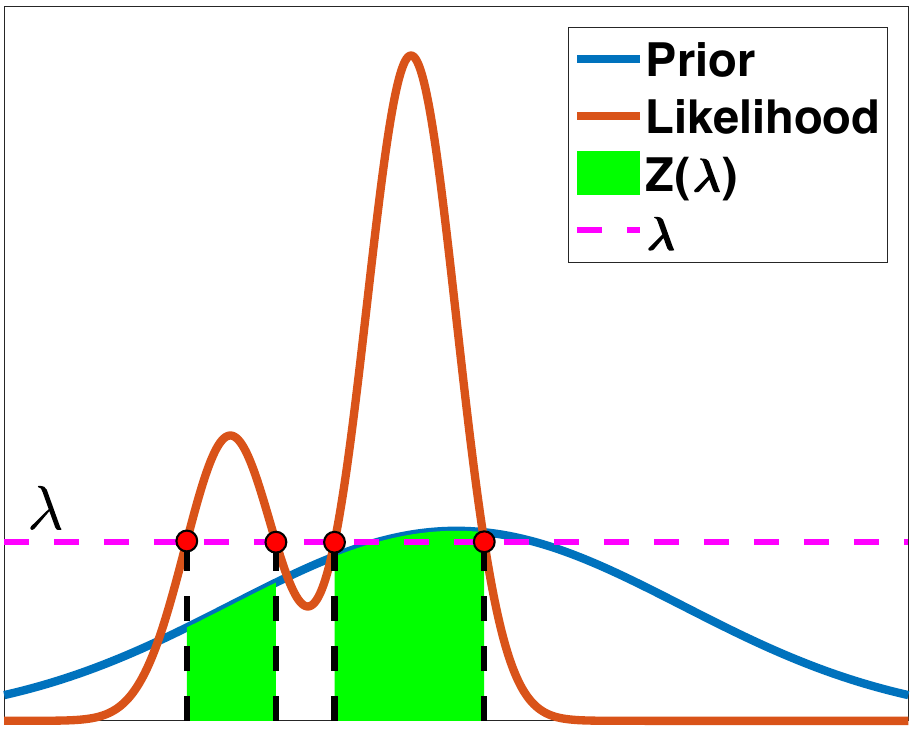}}
\hspace{0.5cm}
\subfigure[\label{figNestedb}]{\includegraphics[width=7cm]{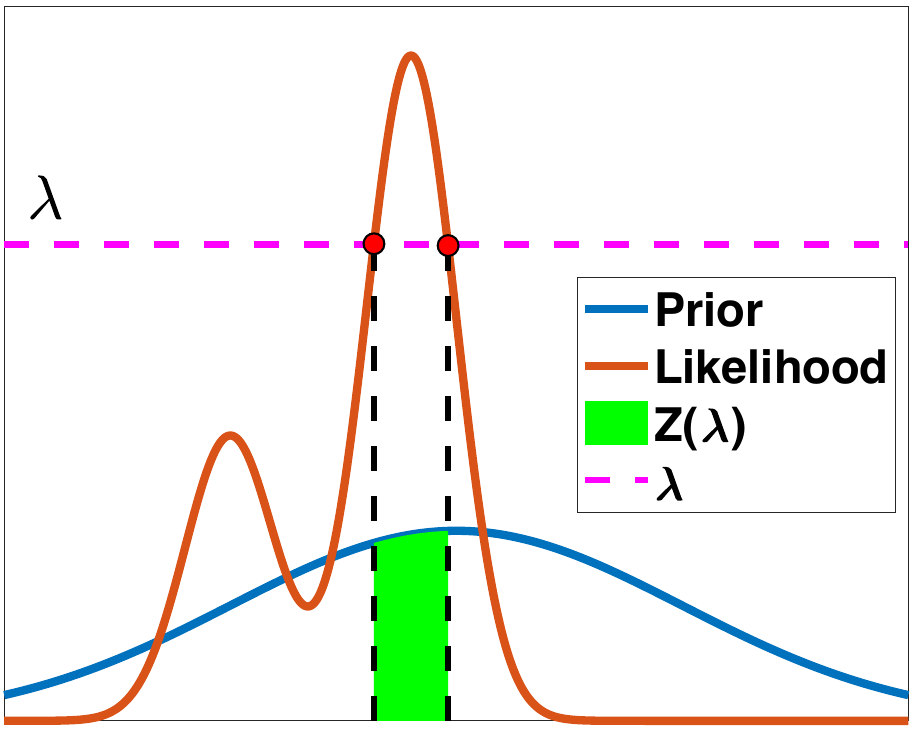}}
\vspace{-0.4cm}
	\caption{ Two examples of the area below {the {\it non-normalized} 
	 truncated prior $\mathbb{I}\{\ell(\y|\x)>\lambda\} g(\x)$. The green area represents the function $Z(\lambda)$.  The normalized truncated prior $g(\x|\lambda)=\frac{1}{Z(\lambda)} \mathbb{I}\{\ell(\y|\x)>\lambda\} g(\x)$ is given in Eq. \eqref{NormTrucatedPriorEq}.  }
	   The red line depicts the likelihood function, and the blue line shows the complete prior density $g(\x)$. The portions of the blue line that define the green areas represent the {the non-normalized 
	 truncated prior $\mathbb{I}\{\ell(\y|\x)>\lambda\} g(\x)$; for obtaining the normalized version $g(\x|\lambda)$ we have to multiply by $1/Z(\lambda)$}.
	Note that in figure (b) the value of $\lambda$ is greater than in figure (a), so that the area $Z(\lambda)$  decreases with respect to Figure (a). Clearly, if we  assume a   $\lambda$ value  bigger than the maximum of the likelihood, then we would have $Z(\lambda)=0$. 
	}
	\label{figNested}
\end{figure}

Note that $Z(\lambda)$ represents also the normalizing constant of the following truncated prior pdf, i.e.,
\begin{gather}\label{NormTrucatedPriorEq}
\boxed{
g(\x|\lambda)=\frac{1}{Z(\lambda)}\mathbb{I}\{\ell(\y|\x)>\lambda\} g(\x)=
\left\{
\begin{split}
&{ \frac{g(\x)}{Z(\lambda)}}, \quad \mbox{ if } \ell(\y|\x)> \lambda,  \\
& 0,  \qquad \mbox{ otherwise, }
\end{split}
\right. \qquad \mbox{{with $0<\lambda <\lambda_{\max}$},}}
\end{gather}
 { where $\lambda_{\max}=\sup \ell(\y |\x)$.
 Since we have assumed $\Theta$ compact, $\ell$ continuous, and assuming now that the likelihood admits a unique global maximum, we can define the maximum likelihood (ML) estimator as $\widehat{\x}_{\text{ML}}=\arg\max \ell({\bf y}|\x)$,  the pdf $g(\x|\lambda)$ becomes increasingly concentrated around $\widehat{\theta}_{\mathrm{ML}}$ as  $\lambda$ approaches $\lambda_{\max}$.
 More precisely, under the above regularity assumptions, the probability measures induced by $g(\x|\lambda)$ converge weakly to the Dirac probability measure concentrated at the ML estimator, i.e.,
  $ g(\x|\lambda)\overset{w}{\longrightarrow} \delta(\x-\widehat{\x}_{\text{ML}})$ as  $\lambda$ approaches $\lambda_{\max}$ from below,
where $ \delta(\x-\widehat{\x}_{\text{ML}})$   denotes the Dirac probability measure concentrated at the atom $\widehat{\x}_{\mathrm{ML}}$, assigning probability one to $\widehat{\x}_{\mathrm{ML}}$. Hence, with the previous assumptions and recalling that we are assuming $\ell(\y |\x)>0$ for all $ \x \in \Theta$ (so that $Z(0)=1$ and $\mathbb{I}\{\ell(\y|\x)>0\}=1$),} we have the two extreme cases:
\begin{align}
& g(\x|0)=g(\x), \\
 &g(\x|\lambda) \overset{w}{\longrightarrow} \delta(\x-\widehat{\x}_{\text{ML}}) \quad \mbox{ with } \quad {\lambda \uparrow  \lambda_{\texttt{max}} = \ell({\bf y}|\widehat{\x}_{\text{ML}})},
\end{align}
{where $\lambda \uparrow \lambda_{\texttt{max}}$ means  that $\lambda$ approaches $\lambda_{\max}$ from below.} Two graphical examples of $g(\x|\lambda)$ and $Z(\lambda)$ are given in Figure \ref{figNested}.  The red line depicts the likelihood function, and the blue line shows the complete prior density $g(\x)$. The portions of the blue line that define the green areas represent { the {\it unnormalized} truncated prior, i.e.,  $\mathbb{I}\{\ell(\y|\x)>\lambda\} g(\x)$. To normalize it, we have to divide by the value $Z(\lambda)$ (that is represented by the green area), i.e., obtaining $g(\x|\lambda)=\frac{1}{Z(\lambda)}\mathbb{I}\{\ell(\y|\x)>\lambda\} g(\x)$.}

{\rem It is important to emphasize that the truncation of the prior density is defined through the likelihood function, as shown in Figure \ref{figNested}. In particular, determining the support of the truncated prior - that is, the region where $g(\x|\lambda)>0$ is well defined and strictly positive - would in general require inverting the likelihood function (which is generally impossible or computationally complex).   }

\subsection{Sampling  from the truncated prior $g(\cdot|\lambda)$}  

Given a fixed value {$0<\lambda_0 <\lambda_{\max}$}, in order to  generate samples from $g(\x|\lambda_0)$  one possibility is to use an MCMC procedure.
However, as an example, the following acceptance-rejection procedure can be also employed \cite{martino2018independent}:
\newline
\fbox{%
\begin{minipage}{0.98\linewidth}
\begin{enumerate}
\item For $n=1,...,N$:
\begin{enumerate}
\item Draw $\x'\sim g(\x)$.   
\item if $\ell(\y|\x')> \lambda_0$ then set $\x_{n|0}=\x'$ and $\lambda_{n|0}=\ell(\y|\x')$.
\item if $\ell(\y|\x')\leq \lambda_0$, then reject $\x'$ and repeat from step 1(a).
\end{enumerate}
\item Return $\{\x_{n|0}\}_{n=1}^N$ and $\{\lambda_{n|0}\}_{n=1}^N$, where all $\x_{n|0} \sim g(\x|\lambda_0)$ and all $\lambda_{n|0} \sim F(\lambda|\lambda_0)$, given below.
\end{enumerate}
\end{minipage}}
\newline
\newline
Observe that $\x_{n|0} \sim g(\x|\lambda_0)$, for all $n=1,...,N$, and the probability of accepting a generated sample $\x'$ is exactly {$Z(\lambda_0)=\mathbb{P}\left(\ell(\y|\x)>\lambda_0\right)$ with $\x \sim g(\x)$, by definition.}  The values $\lambda_{n|0}=\ell(\y|\x_{n|0})$ where $\x_{n|0} \sim g(\x|\lambda_0)$, have the following {\it truncated} cumulative function:
\begin{equation}
F(\lambda|\lambda_0)={\mathbb{P}\Big(\ell(\y|\x)\leq \lambda \big|\ell(\y|\x)> \lambda_0\Big)}=\frac{F(\lambda)-F(\lambda_0)}{1-F(\lambda_0)},
 \quad \mbox{ with } \lambda\geq \lambda_0, \mbox{ and }  \x\sim g(\x|\lambda_0),
\end{equation}
i.e., we can write $\lambda_n\sim F(\lambda|\lambda_0)$. { Recall that $F(\lambda)=\mathbb{P}\left(\ell(\y|\x)\leq\lambda\right)=\mathbb{P}\left(L\leq\lambda\right)$, with  $\x \sim g(\x)$ and we have assumed the continuity of distribution of $L$. Hence,} considering the steps above, (a) draw $\x'\sim g(\x)$ and (b) if $\ell(\y|\x')> \lambda_0$ then set $\lambda_n=\ell(\y|\x')$ (otherwise reject $\x'$ and repeat), we can generate samples according to the truncated cumulative function $F(\lambda|\lambda_0)$. Setting $\lambda_0=0$, we have $F(\lambda|0)=F(\lambda)$ that is the cumulative function defined in Eq. \eqref{EqF}. Recall that $F(0)=0$ and $F(\lambda')=1$ for all  $\lambda' \geq {\sup \ell(\y|\x)}$. Similarly, we have $F(\lambda'|\lambda_0)=0$ for all $\lambda'< \lambda_0$ and $F(\lambda'|\lambda_0)=1$ for all  $\lambda' \geq {\sup \ell(\y|\x)}$.

{
{\rem\label{SuperIMPRem} Consider the accept-reject procedure described above and let $\x_1,\ldots,\x_N$ be the $N$ accepted samples, drawn independently from $g(\x)$ and satisfying $\ell(\y|\x_n)>\lambda_0$ for $n=1,\ldots,N$. Consequently, the population $\mathcal{P}=\{\x_n\}_{n=1}^N$ consists of independent samples from $g(\x|\lambda_0)$. Now, let $\lambda'>\lambda_0$ be a fixed, nonrandom threshold and consider the event that all $N$ samples satisfy the stricter constraint $\ell(\y|\x_n)>\lambda'$. Conditional on this event, the samples remain independent and are distributed according to $g(\x|\lambda')$.}

{\rem\label{SuperIMPRem2} A key observation is the following. Consider again the $N$ accepted samples $\x_1,\ldots,\x_N$, drawn independently from $g(\x)$ and satisfying $\ell(\y|\x_n)>\lambda_0$ for all $n=1,\ldots,N$. Define $\lambda_{\texttt{min}}=\min\limits_{n=1,\ldots,N}\ell(\y|\x_n)$ and let $\x_{n'}$ denote the sample attaining this minimum, so that $\lambda_{\texttt{min}}=\ell(\y|\x_{n'})>\lambda_0$. After removing $\x_{n'}$, the remaining $N-1$ samples can be reused as independent draws from $g(\x|\lambda_{\texttt{min}})$, conditionally on $\lambda_{\texttt{min}}$ (see also Figure \ref{figNested2}), provided that the likelihood values have a continuous distribution, so that ties occur with probability zero. 
}
}

\subsection{Distribution of $a_{n|0}=Z(\lambda_{n|0})$ with $\lambda_{n|0} \sim F(\lambda|\lambda_0)$}
\label{ZandSampling3}

Recall that $Z(\lambda)$ is non-increasing, then $Z(\lambda)\leq Z(\lambda_0)$ if $\lambda_0\leq \lambda$. Moreover, recall that with $Z(0)=1$ and $Z(\lambda')=0$ for all  $\lambda' \geq {\sup \ell(\y|\x)}$. With similar arguments of Section \ref{SectquiZ} and
similarly to Eq. \eqref{AquiAn2}, we can write:
\begin{equation}\label{AquiAn2_otra}
\boxed{
\begin{aligned} 
 \mbox{ if }  & \quad \x_{n|0} \sim g(\x|\lambda_0),   \\ 
 \mbox{ and } &\quad  \lambda_{n|0}=\ell(\y|\x_{n|0}) \sim F(\lambda|\lambda_0), \\
  \quad 
  \mbox{ then }  &\quad  a_{n|0}=Z(\lambda_{n|0}) \sim \mathcal{U}([0,a_0]),  \quad \mbox{ with }a_0=Z(\lambda_0)\leq 1.  
\end{aligned}
}
\end{equation}
Clearly, the expression above ensured that $a_{n|0}=Z(\lambda_n)\leq a_0= Z(\lambda_0)$,  i.e., $a_{n|0} \leq a_0$ (as expected since $Z(\lambda)$ is non-increasing and $\lambda_n\geq \lambda_0$).
 Since $a_{n|0} \sim \mathcal{U}([0,a_0])$, we can also write:
\begin{align}
\fbox{$\displaystyle\widetilde{a}_n=\dfrac{a_{n|0}}{a_0}{=\dfrac{Z(\lambda_{n|0})}{Z(\lambda_0)}} \sim \mathcal{U}([0,1]), \quad \forall n=1,...,N.$} 
\end{align}
{{\rem If we can draw samples $\x_{n|0}$ exactly from $g(\x|\lambda_0)$ and evaluating $ \lambda_{n|0}=\ell(\y|\x_{n|0})$, the ratio of areas $\frac{a_{n|0}}{a_0}=\frac{Z(\lambda_{n|0})}{Z(\lambda_0)}$ is uniformly distributed on $[0,1]$ for every $\lambda_0\in (0,\lambda_{\max})$. This is a powerful result, as it holds regardless of the specific forms of the likelihood and prior functions and for any (admissible) values of $\lambda_0$ and $\lambda_{n|0}$. 
This theoretical result, which relies on the ability to sample exactly from $g(\x|\lambda_0)$, is a key property of the NS procedure and provides the foundation for the subsequent theoretical developments.}}

\subsection{Distribution of $\widetilde{a}_{\texttt{max}}$}
\label{ZandSampling4}
 Let us consider $\lambda_{1|0},....,\lambda_{N|0} \sim F(\lambda|\lambda_0)$ and  the minimum and maximum values   
\begin{equation}\label{IMP_amax0}
\lambda_{\texttt{min}|0}=\min_n \lambda_{n|0}, \quad a_{\texttt{max}|0}=Z(\lambda_{\texttt{min}|0}), \quad  \mbox{ and } \quad \widetilde{a}_{\texttt{max}}=\frac{a_{\texttt{max}|0}}{a_0}=\frac{Z(\lambda_{\texttt{min}|0})}{Z(\lambda_0)}.
\end{equation}
Let us recall $\widetilde{a}_n=\frac{a_{n|0}}{a_0} \sim \mathcal{U}([0,1])$.
Then, note that $\widetilde{a}_{\texttt{max}}$ is  maximum of $N$ uniform random variables 
$$
\widetilde{a}_1, ...,\widetilde{a}_N  \sim \mathcal{U}([0,1]).
$$
Then  it is well-known that the cumulative distribution of the maximum value  
$$
\widetilde{a}_{\texttt{max}}= \max_n \widetilde{a}_n \sim \mathcal{B}(N,1),
$$
is distributed according to a Beta distribution $\mathcal{B}(N,1)$, i.e., with cumulative function
$F_{\texttt{max}}(\widetilde{a})= \widetilde{a}^N$ and density $f_{\texttt{max}}(\widetilde{a})=\frac{dF_\text{max}(\widetilde{a})}{d\widetilde{a}}= N\widetilde{a}^{N-1}$ \cite[Section 2.3.6]{martino2018independent}. In summary, we have
\begin{equation}\label{IMP_amax}
\boxed{
\begin{aligned}
	&\widetilde{a}_{\texttt{max}}= \frac{Z(\lambda_{\texttt{min}|0})}{Z(\lambda_0)} \sim \mathcal{B}(N,1),  \\
	\mbox{ where } \quad &\lambda_{\texttt{min}|0}=\min_n \lambda_{n|0}, \\
 \text{ and } \quad &\lambda_{1|0},...,\lambda_{N|0} \sim F(\lambda|\lambda_0).  
\end{aligned}
}
\end{equation}
{ It is important to remark that all these expressions from \eqref{AquiAn2_otra} to \eqref{IMP_amax} rely on the continuity of the distribution of the random variable $L=\ell(\y|\x)$ with $\x \sim g(\x)$.}
 This result is important for deriving the standard version of the  nested sampling method, described in the next section. A summary of the relationships presented above is provided in Table \ref{SummaryVariables}.

\begin{table}[!h]
	\caption{Summary of the relationships among some random variables introduced above.}
	\label{SummaryVariables}
	\begin{center}
		\begin{tabular}{|c||c|}  
		   	\hline 
		     {\bf Sections} & {\bf Relationships} \\
			\hline 
			\hline 
\hline
  & \\		      
\multirow{4}{* }{\ref{ZandSampling3}}&$\mbox{ If } \x_{n|0} \sim g(\x|\lambda_0), \quad \mbox{ and  }  \quad \lambda_{n|0}=\ell(\y|\x_{n|0}) \sim F(\lambda|\lambda_0),$ \\
& \\
& $\widetilde{a}_n=\dfrac{a_{n|0}}{a_0}=\dfrac{Z(\lambda_{n|0})}{Z(\lambda_0)} \sim \mathcal{U}([0,1]).$ \\
 &\\
 \hline
\multirow{8}{*}{\ref{ZandSampling4}}  &\\
 &$\mbox{ If } \x_{n|0} \sim g(\x|\lambda_0), \mbox{ and  }  \lambda_{n|0}=\ell(\y|\x_{n|0}) \sim F(\lambda|\lambda_0), \quad n=1,...,N,$ \\
 & \\ 
  & $ \lambda_{\texttt{min}|0}=\min_n \lambda_{n|0},$	\\ 
&	\\
& $\widetilde{a}_{\texttt{max}}=\max \widetilde{a}_n=  \dfrac{Z(\lambda_{\texttt{min}|0})}{Z(\lambda_0)}  \sim \mathcal{B}(N,1)$.	\\
& \\
			\hline
		\end{tabular}
	\end{center}
	
\end{table}

\section{A detailed description of  nested sampling (NS) }\label{NestedSampling}

\subsection{Form of the NS estimator}
\label{QuadratureNSsect}

\begin{figure}[!h]
	\centering
\includegraphics[width=8.5cm]{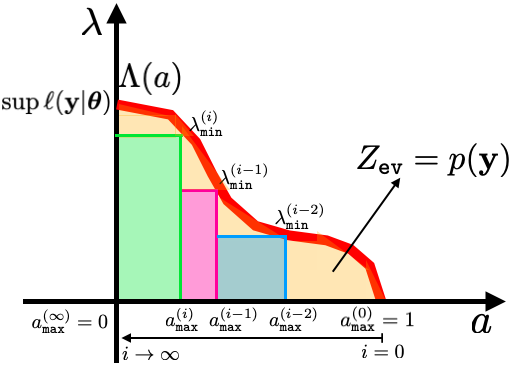}
	\caption{{Graphical representation of the adaptive quadrature scheme provided by the NS procedure, where the nodes $(a_{\texttt{max}}^{(i)},\lambda_{\texttt{min}}^{(i)})$, are automatically chosen by the algorithm. More precisely, in practice, the algorithm first chooses the value $\lambda_{\texttt{min}}^{(i)}= \Lambda(a_{\texttt{max}}^{(i)})$ and then estimate/approximate the corresponding value $ a_{\texttt{max}}^{(i)}$, with $\widehat{a}_i \approx a_{\texttt{max}}^{(i)}$ as shown in Eq. \eqref{approxMeanBeta2}.}
	}
	\label{figNested_otra_quadrature}
\end{figure}

Nested sampling (NS) is a technique for estimating the marginal likelihood that exploits the second identity in \eqref{ZasOneDimIntegral} \cite{skilling2006nested, chopin2010properties, polson2014vertical}. NS estimates $Z_{\texttt{ev}}$ by a quadrature using nodes (in {\it decreasing} order),
$$
0<a_{\texttt{max}}^{(I)}<\dots<a_{\texttt{max}}^{(1)}<a_{\texttt{max}}^{(0)}=1,
$$
and the {rectangular} quadrature formula:
\begin{align}
\label{EqNestedZ}
\fbox{$\displaystyle	\widehat{Z}_{\texttt{ev}} =  \sum_{i=1}^{I} (a_{\texttt{max}}^{(i-1)} -a_{\texttt{max}}^{(i)} )\Lambda(a_{\texttt{max}}^{(i)})=  \sum_{i=1}^{I}  (a_{\texttt{max}}^{(i-1)} -a_{\texttt{max}}^{(i)} ) \lambda_{\texttt{min}}^{(i)},$}
\end{align}
with $a_{\texttt{max}}^{(0)}=1$ {and recall that we always have $a_{\texttt{max}}^{(i-1)} -a_{\texttt{max}}^{(i)}>0$}. Furthermore, we will see that the NS construction yields that
\begin{align}
\label{EqIalmostInf}
\fbox{$\displaystyle \lim_{I \rightarrow \infty} a_{\texttt{max}}^{(I)}=a_{\texttt{max}}^{(\infty)}=0.$}
\end{align}
See  Figure \ref{figNested_otra_quadrature} for a graphical representation.

{\rem Recall  that $\lambda=\ell(\y|\x)$'s represent likelihood values, and $a$'s represent  normalized area values, hence contained in $[0,1]$. }
\newline
In a standard quadrature rule, we should to specify the grid points  $a_{\texttt{max}}^{(i)}$'s (possibly well-located, with a suitable strategy) and the corresponding values $\lambda_{\texttt{min}}^{(i)}= \Lambda(a_{\texttt{max}}^{(i)})$. {
 However,
 the NS algorithm automatically selects the values  $\lambda_{\texttt{min}}^{(i)}$ and approximates the corresponding values $a_{\texttt{max}}^{(i)}$.} Recall that the function $\Lambda(a)$, and its inverse $a=\Lambda^{-1}(\lambda)=Z(\lambda)$, are generally intractable and unknown, so that  it is not even possible to evaluate $\Lambda(a)$ at a grid of chosen $a_{\texttt{max}}^{(i)}$'s. 

{\rem\label{REM-AutomaticQuad} {The quadrature values $a_{\texttt{max}}^{(i)}$, $\lambda_{\texttt{min}}^{(i)}$  in Eq. \eqref{EqNestedZ}, are automatically chosen by the NS procedure (see Figure \ref{figNested_otra_quadrature}).} The NS algorithm first suitably selects the ordinates $\lambda_{\texttt{min}}^{(i)}$ (likelihood values) {with a stochastic process that accumulates more values close to $\sup \ell(\y|\x)$,} and then uses  approximations $\widehat{a}_i$ of the corresponding (unknown) values $a_{\texttt{max}}^{(i)}=Z(\lambda_{\texttt{min}}^{(i)})$. This last step is possible since the distribution of $a_{\texttt{max}}^{(i)}$ is known under the given assumptions (see Section \ref{ZandSampling4}).}

\noindent
 {Clearly, alternative quadrature rules could be also employed. For instance, the trapezoidal quadrature formula for NS is: 
$$
\widehat{Z}_{\texttt{ev}}=\sum_{i=1}^I \frac{\Lambda(a_{\texttt{max}}^{(i-1)})+\Lambda(a_{\texttt{max}}^{(i)})}{2}\left(a_{\max}^{(i-1)}-a_{\max}^{(i)}\right).
$$
Defining $\gamma_i = a_{\max}^{(i-1)}-a_{\max}^{(i)}$, in each interval $[a_{\max}^{(i-1)},a_{\max}^{(i)}]$, bounds for the quadrature error of the rectangular and trapezoidal formulas can be found:
\begin{align}\label{ErrQuadEq}
\left|E_\texttt{rect}^{(i)}\right| \leq \frac{1}{2} \sup \left|\Lambda^{\prime}(\xi)\right|\gamma_i^2, \qquad \left|E_\texttt{trap}^{(i)}\right| \leq \frac{1}{12} \sup \left|\Lambda^{\prime \prime}(\xi)\right|\gamma_i^3,
\end{align}
where $\xi \in [a_{\texttt{max}}^{(i)}, a_{\texttt{max}}^{(i-1)}]$ and $\Lambda^{\prime}(\xi)$, $\Lambda^{\prime\prime}(\xi)$ denote the first and second derivatives of $\lambda(a)$, respectively, evaluated at $\xi$. However, in the NS procedure, the prior areas $a_{\texttt{max}}^{(i)}$ are estimated, and the likelihood levels $\lambda_{\texttt{min}}^{(i)}=\Lambda(a_{\texttt{max}}^{(i)})$ are random order statistics. Consequently, the error formulas above describe only a partial error of the NS algorithm (i.e., only the quadrature component). The complete NS error of the evidence estimator must include the statistical error components of the overall procedure. See Section \ref{StocErrSect}, for some other details.
}


\subsection{Choice of $\lambda_{\texttt{min}}^{(i)}$ and $a_{\texttt{max}}^{(i)}$ in nested sampling}
\label{superIMPNested}

\begin{figure}[!h]
{
	\centering
\subfigure[]{\includegraphics[width=7.5cm]{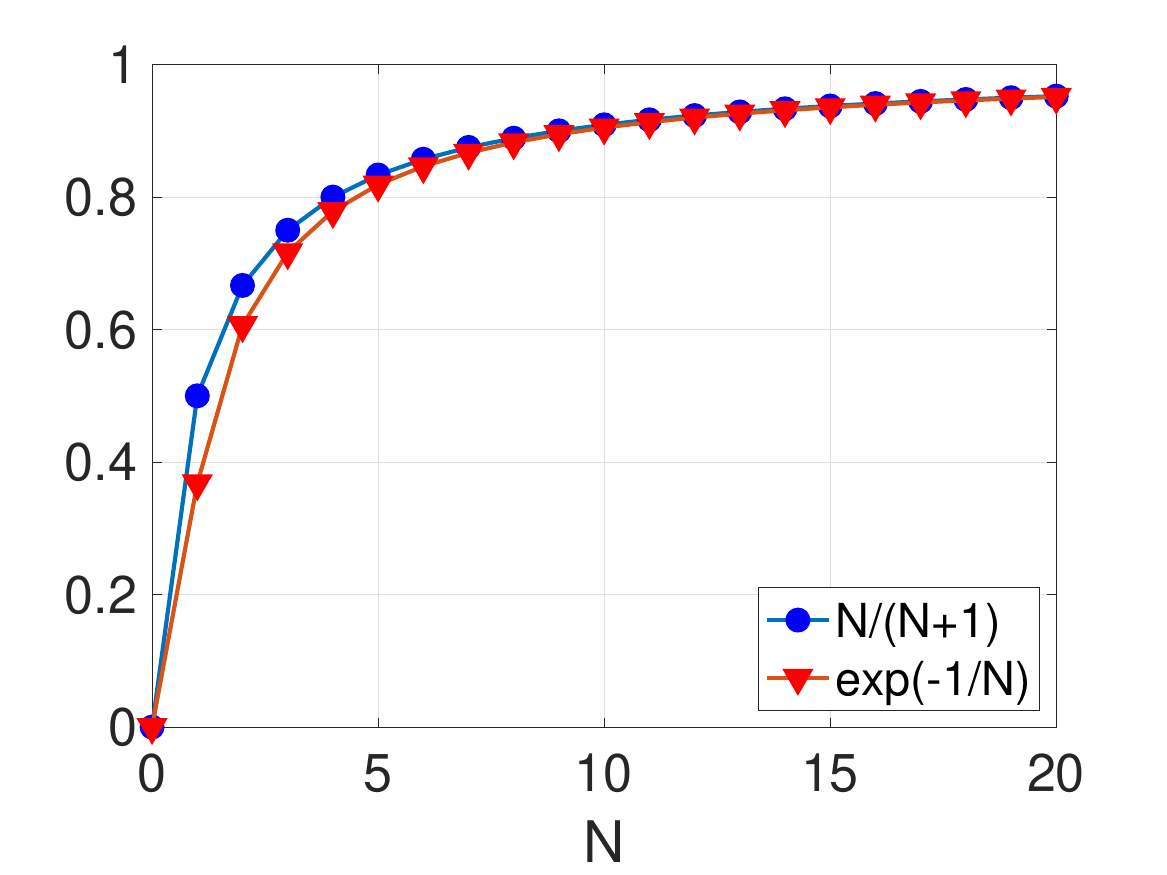}}
\hspace{0.5cm}
\subfigure[]{\includegraphics[width=7.5cm]{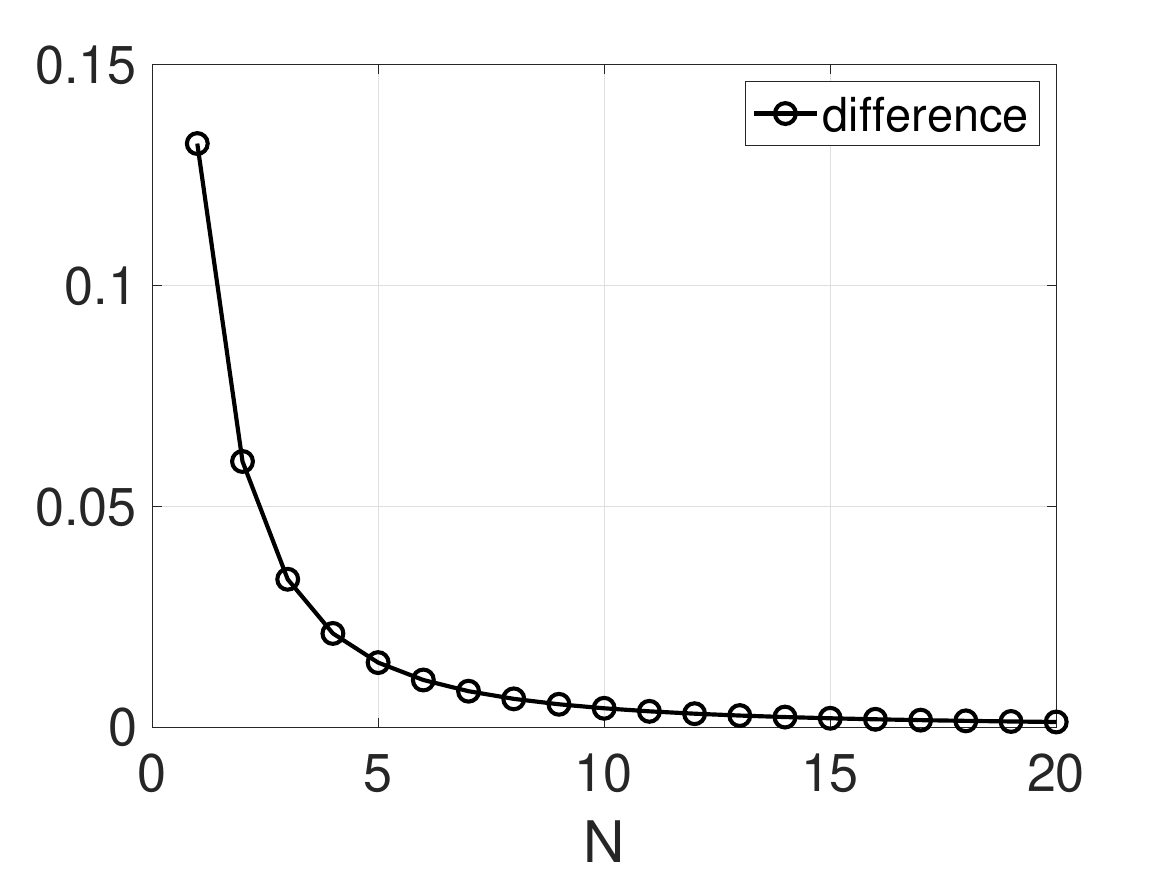}}
\vspace{-0.4cm}
	\caption{{\bf (a)}The two functions $\frac{N}{N+1}$ and $\exp\left(-\frac{1}{N}\right)$ as function of the integer $N$. {\bf (b)} The difference between the two functions (versus $N$) that converges to zero, thereby justifying Eq. \eqref{approxMeanBeta}.
	}
	\label{DiffFig}
	}
\end{figure}

In this section, we adopt a slightly simplified notation compared to the previous part of the work, as nested sampling involves additional indices that may otherwise obscure the exposition and hinder the reader's understanding.
Nested sampling employs an iterative procedure  in order to  generate an {\it increasing} sequence of likelihood ordinates $\lambda_{\texttt{min}}^{(i)}$, $i=1,...,I$, such that
\begin{equation}
\lambda_{\texttt{min}}^{(1)}<\lambda_{\texttt{min}}^{(2)}<\lambda_{\texttt{min}}^{(3)}....<\lambda_{\texttt{min}}^{(I)}.
\end{equation}
Table \ref{tableNestedSampling} provides a compact and complete description of the standard NS algorithm that is based on the sampling of the truncated prior pdf $g(\x|\lambda_{\texttt{min}}^{(i-1)})$ (see Section \ref{ZandSampling2} for the related details), where $i$ denotes an iteration index. 
The nested sampling approach is explained with more details below:
\begin{itemize}
\item At the first iteration ($i=1$), we set $\lambda_{\texttt{min}}^{(0)}=0$ and $a_{\texttt{max}}^{(0)}=Z(\lambda_{\texttt{min}}^{(0)})=1$. Then, $N$ samples are drawn from the prior $\x_n\sim g(\x|\lambda_{\texttt{min}}^{(0)})=g(\x)$ obtaining an (initial) cloud 
\begin{align}
\mathcal{P}_0=\{\x_1,...,\x_N\}=\{\x_n\}_{n=1}^N,
\end{align}
often called {\it set of ``live points''} and  then set $\lambda_n=\ell(\y|\x_n)$, i.e.,
$\{\lambda_n\}_{n=1}^N\sim F(\lambda)$ as shown in Section \ref{ZandSampling}. Thus, the first ordinate is chosen as
$$
\lambda_{\texttt{min}}^{(1)}= \min_{n}  \lambda_n= \min\limits_{\x\in \mathcal{P}_0}\ell(\y|\x)=\ell(\y|\x_{\texttt{rem}}^{(1)}), 
$$
where $\x_{\texttt{rem}}^{(1)}=\arg\min\limits_{\x\in \mathcal{P}_0}\ell(\y|\x)$ is also removed from $\mathcal{P}_0$, i.e., $\mathcal{P}_1=\mathcal{P}_0\backslash \{\x_{\texttt{rem}}^{(1)}\}$ (at this stage, now we have $|\mathcal{P}_1|=N-1$). Moreover, since $\{\lambda_n\}_{n=1}^N\sim F(\lambda)$,  using the result in  Eq. \eqref{IMP_amax}, we have that 
$$
\widetilde{a}_{\texttt{max}}^{(1)}=\frac{a_{\texttt{max}}^{(1)}}{a_{\texttt{max}}^{(0)}}=\frac{Z(\lambda_{\texttt{min}}^{(1)})}{Z(\lambda_{\texttt{min}}^{(0)})}\sim \mathcal{B}(N,1).
$$
Since $a_{\texttt{max}}^{(0)}=Z(\lambda_{\texttt{min}}^{(0)})=1$, then $\widetilde{a}_{\texttt{max}}^{(1)}=a_{\texttt{max}}^{(1)} \sim \mathcal{B}(N,1)$. 
\item At a generic $i$-th iteration, with $i\geq 2$, a unique additional sample $\x'$ is drawn from the truncated prior $g(\x|\lambda_{\texttt{min}}^{(i-1)})$ and  added to the current cloud of samples, i.e., $\mathcal{P}_{i-1}=\mathcal{P}_{i-1}\cup \x'$ (now we recover $|\mathcal{P}_{i-1}|=N$). First of all, note that the {new} value $\lambda' =\ell(\y|\x')$ is distributed as $F(\lambda|\lambda_{\texttt{min}}^{(i-1)})$ {by construction} (see Section \ref{ZandSampling2}). More precisely, all the $N$ ordinate (likelihood) values: 
$$
\{\lambda_n\}_{n=1}^N=\ell(\y|\x\in\mathcal{P}_{i-1})=\{\lambda_n=\ell(\y|\x_n)  \mbox{ }  \mbox{ for all }  \mbox{ }  \x_n\in \mathcal{P}_{i-1}\},
$$
are distributed as $F(\lambda|\lambda_{\texttt{min}}^{(i-1)})$, i.e., $\{\lambda_n\}_{n=1}^N \sim F(\lambda|\lambda_{\texttt{min}}^{(i-1)})$. This is due to how the population $\mathcal{P}_{i-1}$ has been built in the previous iterations. {For the same reason, at this stage, all the $N$ samples $\x_n\in \mathcal{P}_{i-1}$ are distributed as $g(\x|\lambda_{\texttt{min}}^{(i-1)})$, as stated in Remarks \ref{SuperIMPRem} and \ref{SuperIMPRem2}.}  Then, we compute again the new minimum value as
$$
\lambda_{\texttt{min}}^{(i)}= \min_{n}  \lambda_n=  \min_{\x\in \mathcal{P}_{i-1}} \ell(\y|\x)=\ell(\y|\x_{\texttt{rem}}^{(i)}),
$$
and remove the corresponding sample $\x_{\texttt{rem}}^{(i)}=\arg\min\limits_{\x\in \mathcal{P}_{i-1}}\ell(\y|\x)$, i.e., we set $\mathcal{P}_i=\mathcal{P}_{i-1} \backslash \{\x_{\texttt{rem}}^{(i)}\}$ and the procedure is repeated.
 Moreover, since $\lambda_{\texttt{min}}^{(i)}$ is the minimum value of $\{\lambda_1,...,\lambda_N\}\sim  F(\lambda|\lambda_{\texttt{min}}^{(i-1)})$, in Section \ref{ZandSampling4} we have seen that  
\begin{equation}\label{ESTAAQUI_INCREIBLE}
\widetilde{a}_{\texttt{max}}^{(i)}=\frac{a_{\texttt{max}}^{(i)}}{a_{\texttt{max}}^{(i-1)}}=\frac{Z(\lambda_{\texttt{min}}^{(i)})}{Z(\lambda_{\texttt{min}}^{(i-1)})} \sim \mathcal{B}(N,1),
\end{equation}
where we have used Eq. \eqref{IMP_amax}.  Note that we have also found the recursion among the following random variables,
\begin{equation}\label{RecursionAi}
a_{\texttt{max}}^{(i)}= \widetilde{a}_{\texttt{max}}^{(i)} a_{\texttt{max}}^{(i-1)}, 
\end{equation}
for  $i=1,...,I$, recalling that $a_{\texttt{max}}^{(0)}=1$. 
\item { We do not know the actual values of all the areas $a_{\texttt{max}}^{(i)}$.  The idea is to estimate the random values $a_{\texttt{max}}^{(i)}$ (and $\widetilde{a}_{\texttt{max}}^{(i)}$) for all $i$, by their expected values. With this purpose, we have to observe that:}
\begin{itemize}
\item  The random value $\widetilde{a}_{\texttt{max}}^{(i)}$ could be estimated and replaced with the expected value of the Beta distribution $\mathcal{B}(N,1)$, i.e., $\widetilde{a}_{\texttt{max}}^{(i)} \approx  \mathbb{E}[ \widetilde{a}_{\texttt{max}}^{(i)}]$, where
\begin{equation} \label{approxMeanBeta}
 \mathbb{E}[ \widetilde{a}_{\texttt{max}}^{(i)}]  =\mathbb{E}[\mathcal{B}(N,1)]=\frac{N}{N+1} \approx \exp\left(-\frac{1}{N}\right).
\end{equation} 
where $\mathbb{E}[\mathcal{B}(N,1)]=\frac{N}{N+1}$ is used as estimator, and $ \exp\left(-\frac{1}{N}\right)$ becomes a very good approximation of {$\frac{N}{N+1}$ as $N$ grows, as depicted in Figure \ref{DiffFig} (this is already a very good approximation for $N\geq 7$). Another explanation is often given: since $\mathbb{E}[ \log\widetilde{a}_{\texttt{max}}^{(i)}]=-\frac{1}{N}$, if we take then the exponential $\exp\left(\mathbb{E}[ \log\widetilde{a}_{\texttt{max}}^{(i)}]\right)=\exp\left(-\frac{1}{N}\right)$.\footnote{{Clearly, $\exp\left(\mathbb{E}[ \log\widetilde{a}_{\texttt{max}}^{(i)}]\right)\leq \mathbb{E}[\widetilde{a}_{\texttt{max}}^{(i)}]$ by Jensen inequality, i.e., $f(\mathbb{E}[A])\leq \mathbb{E}[f(A)]$ where  $f(A)=\exp(A)$ is a convex function, and $A=\log \widetilde{a}_{\texttt{max}}^{(i)}$ (see also Figure \ref{approxMeanBeta}).}} In any case, the need of this approximation is obscure since $ \mathbb{E}[ \widetilde{a}_{\texttt{max}}^{(i)}]  =\frac{N}{N+1} $ is analytically available, i.e., the use of $\exp(-1/N)$ is not strictly necessary. }  
\item { Additionally we need to assume}
{\it the independence of the live points in the population at $i$-th iteration,} { conditional on $\lambda_{\texttt{min}}^{(i-1)}$.  Under this  assumption, 
    in Eq. \eqref{RecursionAi}, we can use} the property of the expectation valid for independent variables\footnote{{ Assuming the independence of the live points in the population at $i$-th iteration, the variables $\widetilde{a}_{\texttt{max}}^{(i)}$ and $a_{\texttt{max}}^{(i-1)}$ are independent, due to Eq. \eqref{ESTAAQUI_INCREIBLE}, i.e., that ratio has always the same beta distribution, $\mathcal{B}(N,1)$, for all $i$.}}, 
\begin{equation}\label{ProdExpectEQ}
\E[a_{\texttt{max}}^{(i)}]= \E[\widetilde{a}_{\texttt{max}}^{(i)}]  \ \E[a_{\texttt{max}}^{(i-1)}],
\end{equation}
{and also using the approximation in \eqref{approxMeanBeta}, the recursion in Eq. \eqref{RecursionAi} becomes:}
\begin{equation}\label{approxMeanBeta2}
a_{\texttt{max}}^{(i)}\approx \exp\left(-\frac{1}{N}\right) a_{\texttt{max}}^{(i-1)}{=a_{\texttt{max}}^{(0)} \prod_{k=1}^i \exp\left(-\frac{1}{N}\right)}=\exp\left(-\frac{i}{N}\right), \quad  {\forall i,}
\end{equation} 
 {where we have also considered that $a_{\texttt{max}}^{(0)}=1$.}
Then, denoting $\widehat{a}_i=\exp\left(-\dfrac{i}{N}\right)$, we can use $\widehat{a}_i$ as an approximation of $a_{\texttt{max}}^{(i)}$, {that are the coefficients employed in Eq. \eqref{EqNestedZ} and, after the corresponding substitution with $\widehat{a}_i$, in Eq. \eqref{FinalNestedEstEQ}. Clearly, note that $\lim\limits_{i\rightarrow \infty}\widehat{a}_i=0$.}
\end{itemize}
\end{itemize}

\begin{figure}[!h]
	\centering
\includegraphics[width=18cm]{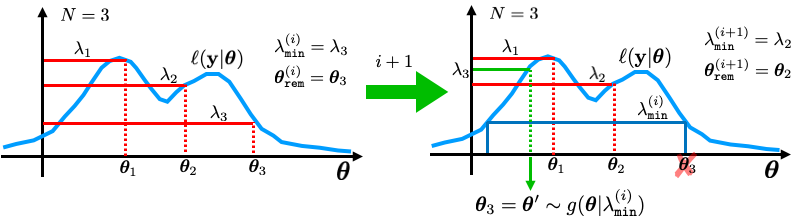}
	\caption{Graphical representation of the NS procedure from the $i$-th iteration  to the $(i+1)$-th iteration, with $N=3$ live points within the population $\mathcal{P}_i=\{\x_1,\x_2,\x_3\}$. The blue line represents the likelihood function $\ell({\bf y}|\x)$. At $i$-th iteration, the point $\x_3$ has the lowest likelihood value, i.e., $\lambda_{\texttt{min}}^{(i)}=\lambda_3=\ell({\bf y}|\x_3)$. Hence, $\x_3$ is removed from the population $\mathcal{P}_i$ and, at the next  $(i+1)$-th iteration, a new sample is generated {$\x_3=\x'\sim g(\x|\lambda_{\texttt{min}}^{(i)})$} and add to the set $\mathcal{P}_i$ in order to keep fixed the number of live points $N=|\mathcal{P}_i|=3$. The new sample generation, at the $(i+1)$-th iteration, is done according to the truncated prior $g(\x|\lambda_{\texttt{min}}^{(i)})$ (see Figure \ref{figNested}). New live points tend to be added in regions of high probability mass (and{, after several iterations,} near to the global mode).
	}
	\label{figNested2}
\end{figure}

 {
{\rem It is worth noting that, when rejection sampling is used to sample from $g(\x|\lambda_{\texttt{min}}^{(i)})$, the quantity $a_{\texttt{max}}^{(i)}=Z(\lambda_{\texttt{min}}^{(i)})$ can be also estimated using the observed acceptance rate, since the theoretical acceptance probability is exactly $Z(\lambda_{\texttt{min}}^{(i)})$, assuming that the proposal distribution is the prior $g(\x)$. }}
\newline
The resulting algorithm is given in Table \ref{tableNestedSampling}.  Figure \ref{figNested2} illustrates, with $N=3$, the removal and replacement steps of the live points within the population $\mathcal{P}$ between the two consecutive iterations. Note that the NS procedure tends to  dynamically add live points near to the main mode or, more generally, to areas of high probability mass. To properly understand the definition and sampling of the truncated prior density, we recommend accurately examining Figure~\ref{figNested}.

{\rem The intuition behind the iterative approach above is to accumulate more ordinates $\lambda_i$ close to the $\sup \ell(\y|\x)$. They are also more dense around $\sup \ell(\y|\x)$. Moreover, using this scheme, we can employ  $\widehat{a}_i=\exp\left(-\frac{i}{N}\right)$ as an approximation of $a_{\texttt{max}}^{(i)}$.}

{\rem An implicit optimization of the likelihood function is performed in the nested sampling algorithm. All  the value $\lambda_n=\ell(\y|\x_n)$ with $\x_n\in \mathcal{P}_i$  approaches the value $\sup \ell(\y|\x)$.
}

{\rem Note that with respect to the $\theta$-space, the NS method is a population sampler where a cloud of points $\mathcal{P}_i=\{\x_n\}_{n=1}^N$ that changes with the iterations, adding and removing one point per iteration.}

 
 \begin{table}[!h]
\caption{\normalsize The standard nested sampling procedure. }
\label{tableNestedSampling}
\begin{tabular}{|p{0.95\columnwidth}|}
   \hline
   \vspace{-0.7cm}
\begin{enumerate}
\item Choose  the number of points per iterations $N$, the total number of iterations $I$ {(big enough, or a stopping rule; see Section \ref{StopRuleSect})}, and set $\widehat{a}_0=1$.
\item Draw $\{\x_{n}\}_{n=1}^N \sim g(\x)$ and define the set $\mathcal{P}_0=\{\x_{n}\}_{n=1}^N$. Let us also define the notation 
\begin{equation}
\ell(\y|\x\in\mathcal{P}_0)=\{\lambda_n=\ell(\y|\x_n)  \mbox{ }  \mbox{ for all }  \mbox{ }  \x_n\in \mathcal{P}_0\},
\end{equation}
\item Set $\lambda_{\texttt{min}}^{(1)} =\min\limits_{\x\in \mathcal{P}_0}\ell(\y|\x)=\ell(\y|\x_{\texttt{rem}}^{(1)})$ where $\x_{\texttt{rem}}^{(1)}=\arg\min\limits_{\x\in \mathcal{P}_0}\ell(\y|\x)$.
\item Set $\mathcal{P}_1=\mathcal{P}_0\backslash \{\x_{\texttt{rem}}^{(1)}\}$, i.e., remove $\x_{\texttt{rem}}^{(1)}$ from $\mathcal{P}_0$.
\item Find an approximation $\widehat{a}_1$ of $a_{\texttt{max}}^{(1)}=Z(\lambda_{\texttt{min}}^{(1)})$. One usual choice is $\widehat{a}_1=\exp\left(-\frac{1}{N}\right)$.
\item For $i=2,..,I:$
\begin{enumerate}
\item Draw $\x' \sim g(\x|\lambda_{\texttt{min}}^{(i-1)} )$ and add to the current cloud of samples, i.e., $\mathcal{P}_{i-1}=\mathcal{P}_{i-1}\cup \x'$. 
\item Set  $\lambda_{\texttt{min}}^{(i)} =\min\limits_{\x\in \mathcal{P}_{i-1}}\ell(\y|\x)=\ell(\y|\x_{\texttt{rem}}^{(i)})$ where  $\x_{\texttt{rem}}^{(i)}=\arg\min\limits_{\x\in \mathcal{P}_{i-1}}\ell(\y|\x)$.
\item Set $\mathcal{P}_{i}=\mathcal{P}_{i-1} \backslash \{\x_{\texttt{rem}}^{(i)}\}$, i.e., remove $\x_{\texttt{rem}}^{(i)}$ from the set of {\it live points} $\mathcal{P}_{i-1}$.
\item Find an approximation $\widehat{a}_i$ of $a_{\texttt{max}}^{(i)}=Z(\lambda_{\texttt{min}}^{(i)})$. One usual choice is 
\begin{equation}
\widehat{a}_i=\exp\left(-\frac{i}{N}\right) \approx a_{\texttt{max}}^{(i)},
\end{equation}
The rationale behind this choice is explained in the section above.
\end{enumerate}
\item Return 
\begin{align}\label{FinalNestedEstEQ}
	\widehat{Z}_{\texttt{ev}} =  \sum_{i=1}^I(\widehat{a}_{i-1}-\widehat{a}_i)\lambda_{\texttt{min}}^{(i)}= \sum_{i=1}^I(e^{-\frac{i-1}{N}}-e^{-\frac{i}{N}})\lambda_{\texttt{min}}^{(i)}.
\end{align}
\end{enumerate} \\ \\
\hline 
\end{tabular}
\end{table} 

{

{\rem\label{RemNoPartition}  Denoting $
\Theta_i=\left\{\x\in \Theta: \ell(\mathbf{y} | \x)>\lambda_{\texttt{min}}^{(i)}\right\}$ with $
\lambda_{\texttt{min}}^{(0)}=0<\lambda_{\texttt{min}}^{(1)}<\cdots<\lambda_{\texttt{min}}^{(I)}$, then the regions $\Theta_i$ are nested, i.e., $
\Theta_I \subseteq \Theta_{I-1} \subseteq \cdots \subseteq \Theta_0=\Theta$.
Thus, the constrained regions overlap, because each new region is contained in the preceding one. Hence, there is not a partition of the parameter space $\Theta$. However, although the constrained regions in $\Theta$-space overlap, NS constructs disjoint shells by taking subsets defined as the difference $\mathcal{S}_i=\Theta_{i-1} \backslash \Theta_i$. Ignoring boundary aspects, these subsets do not overlap, $
\mathcal{S}_i \cap \mathcal{S}_j=\varnothing, \quad i \neq j$. This is done taking into account the rest $\widehat{a}_{i-1}-\widehat{a}_{i}$ 
 (see also Figure \ref{figNested_otra_quadrature}). Each subset corresponds to a range of likelihood values, and the evidence can be decomposed into contributions from these subsets, plus the final remaining region. }

\subsection{Stopping rules suggested in literature}
\label{StopRuleSect}

The nested sampling in Table \ref{tableNestedSampling} is an idealized/asympotic version of the algorithm where the number of iterations $I$ should be enough, i.e., $I \rightarrow \infty$.
A number of iterations $I$ and/or maximum number of likelihood evaluation can be always considered following a computational budget, but they are not a convergence criterion \cite{Higson19,Keskin2023}.  However, given $I$ finite iteration, the {\it residual evidence}, i.e., the amount of evidence that has not been included in final estimator of the NS procedure in Table \ref{tableNestedSampling}, can be estimated:
\begin{align}
\boxed{
Z_{\mathrm{res}}^{(I)}\approx \int_0^{\widehat{a}_I}  \Lambda(a)da=\int_0^{\exp\left(-\frac{I}{N}\right)}  \Lambda(a)da.}
\end{align}
Moreover, as shown in Figure \ref{figNested0}, we have 
$$
 \Lambda(a) \leq \sup_{\x\in \Theta} \ell(\y|\x) \approx L_{\max}^{(I)} =\max_{\x\in \mathcal{P}_I}   \ell(\y|\x_n), \quad  \x_nÊ\in \mathcal{P}_I,
$$
as a consequence, we have: 
\begin{align}
\boxed{
Z_{\mathrm{res}}^{(I)} \approx \widehat{Z}_{\mathrm{res}}^{(I)}=\exp\left(-\frac{I}{N}\right) L_{\max}^{(I)},}
\end{align}
that is a practical approximation of the residual evidence that can be used for designing a stopping rule \cite{feroz2008multimodal,Keeton11,Keskin2023,Speagle2020dynesty}.  The run is stopped when this becomes sufficiently small relative to the accumulated evidence, for example
$$
\frac{\widehat{Z}_{\mathrm{res}}^{(I)} }{\widehat{Z}_{\texttt{ev}}^{(I)}}<\epsilon, \quad \mbox{with, e.g., $\epsilon=0.01$},
$$
or the same criterion but in logarithmic domain, this is essentially the criterion used by \cite{feroz2008multimodal,Speagle2020dynesty}. An alternative idea is to estimate the variance $\widehat{\sigma}_Z$ of $\widehat{Z}_{\texttt{ev}}^{(I)}$, for instance, using parallel runs of the  NS procedures, and considering the stopping rule $\frac{\widehat{\sigma}_Z }{\widehat{Z}_{\texttt{ev}}^{(I)}}<\epsilon$. A simpler stopping criterion can also be employed, based on the stabilization of the evidence estimate:
$$
\left|\log \widehat{Z}_{\texttt{ev}}^{(I)}-\log \widehat{Z}_{\texttt{ev}}^{(I-k)}\right|<\epsilon.
$$
for some number $k$ of consecutive iterations.
This has an intuitive appeal, but there is an important problem: small changes in the current estimate dc not guarantee that the unexplored region contains negligible evidence. Indeed, narrow posterior peaks can be missed or encountered late, so a criterion based only on the stability of successive estimates can be misleading.

\subsection{Refined finite-run NS estimator}
\label{StopRuleSect2}  

The NS estimator in Eq. \eqref{FinalNestedEstEQ} is based on the contributions of the discarded points and is typically considered in the limit $I\to\infty$. For a finite number of iterations, the estimator can be refined by also accounting for the contribution of the remaining live points, yielding a finite-run approximation \cite{Higson19,Keskin2023}.

\subsubsection{Contribution of the remaining live points}

In the same fashion of the residual evidence estimation, we can refine the estimator $\widehat{Z}_{\texttt{ev}}$ in Eq. \eqref{FinalNestedEstEQ}  by incorporating the contribution of the {\it final live points} remaining at the last iteration $I$. Thus, the NS algorithm in Table \ref{tableNestedSampling} stops after $I$ iterations and retains $N$ live points in the population $\mathcal{P}_I$, that we denote here as $\x_1^{(I)}$,..., $\x_N^{(I)}$. The residual area estimated by these live points can be expressed as \cite{higson2019dynamic,Keeton11,Keskin2023,Speagle2020dynesty}:
\begin{align}\label{Eq_LIVE_Z}
\boxed{
\widehat{Z}_{\text{live}}^{(I)}=\widehat{a}_I \frac{1}{N} \sum_{n=1}^N \ell(\mathbf{y} | \x_n^{(I)}).}
\end{align}

\subsubsection{IS interpretation of Eq. \eqref{Eq_LIVE_Z}}

 The estimator above can be seen as an importance sampling (IS) estimator.
Indeed, recall that $a_{\texttt{max}}^{(I)}=Z(\lambda_{\texttt{min}}^{(I)})$ is the prior mass below the unnormalized constrained prior $\mathbb{I}\{\ell(\y|\x)>\lambda_{\texttt{min}}^{(I)}\} g(\x)$. If we 
 draw from the following proposal  density (recall Remarks \ref{SuperIMPRem} and \ref{SuperIMPRem2}):
$$
\x_n^{(I)} \sim g(\x|\lambda_{\texttt{min}}^{(I-1)})=\frac{1}{a_{\texttt{max}}^{(I-1)}} \mathbb{I}\{\ell(\y|\x)>\lambda_{\texttt{min}}^{(I)}\} g(\x), \quad n=1,...,N,
$$ 
the IS estimator of posterior area restricted to the region $\{\x\in \Theta: \mbox{ } \ell(\mathbf{y} | \x)>\lambda_{\texttt{min}}^{(I)} \}$ is given by:
\begin{align}
\widehat{Z}_{\text{IS}}^{(I)}&= \frac{1}{N} \sum_{n=1}^N \frac{\ell(\mathbf{y} | \x_n^{(I)}) g(\x_n^{(I)})}{\frac{1}{a_{\texttt{max}}^{(I)}} \mathbb{I}\{\ell(\y|\x_n^{(I)})>\lambda_{\texttt{min}}^{(I)}\} g(\x_n^{(I)})}, \nonumber \\
&= \frac{1}{N} \sum_{n=1}^N \frac{\ell(\mathbf{y} | \x_n^{(I)})}{\frac{1}{a_{\texttt{max}}^{(I)}}} =\frac{1}{N} \sum_{n=1}^N w_{n}^{(I)},\label{Est_IS_Zlive}\\
&=a_{\texttt{max}}^{(I)} \frac{1}{N} \sum_{n=1}^N \ell(\mathbf{y} | \x_n^{(I)}) \approx  \widehat{a}_I\frac{1}{N} \sum_{n=1}^N \ell(\mathbf{y} | \x_n^{(I)}), \nonumber
\end{align}
that is exactly $\widehat{Z}_{\text{live}}^{(I)}$ above. 

{\rem\label{RefEvidEqREM} Note that we have denoted as $w_{n}^{(I)}=a_{\texttt{max}}^{(I)}\ell(\mathbf{y} | \x_n^{(I)})$  the posterior IS weights of the remaining live points. Indeed, in Eq. \ref{Est_IS_Zlive}, the arithmetic mean of $w_{n}^{(I)}$ provides the area estimation given by the remaining live points at the last iteration $I$.}

\subsubsection{Refined NS estimator}

We can include the additional contribution of the remaining live point to the final NS estimator:
\begin{align}
&\widehat{Z}_{\texttt{ev}}= \sum_{i=1}^I \left(\widehat{a}_{i-1}-\widehat{a}_i\right) \lambda_{\min }^{(i)}+\widehat{Z}_{\text{live}}^{(I)}, \nonumber \\
&\boxed{
\widehat{Z}_{\texttt{ev}}=  \sum_{i=1}^I\left(\widehat{a}_{i-1}-\widehat{a}_i\right) \lambda_{\min }^{(i)} +\widehat{a}_I \frac{1}{N} \sum_{n=1}^N \ell\left(\mathbf{y} |\boldsymbol{\theta}_n^{(I)}\right).} \label{RefEvidEq}
\end{align}

\subsection{Main sources of uncertainty in NS}
\label{StocErrSect}

The quadrature rule considered in Section \ref{QuadratureNSsect} introduces a first error component in the final NS estimator. Here, we describe other  sources of uncertainty in the NS procedure \cite{chopin2010properties,Higson2018,Keeton11,skilling2006nested}. A complete list of the main error components in the NS  algorithm is given below:
\begin{itemize}
\item {\bf Numerical quadrature}: due to the analytic form of the NS estimator, that is discussed in Section \ref{QuadratureNSsect};
\item {\bf Constrained-prior mass estimations:} due to the estimation of  constrained prior areas $a_{\max}^{(i)}$, which are random realizations. This is discussed in Section \ref{Source2Sect} below.
\item {\bf Ordinary Monte Carlo variability:} this component is generated to the random variation in likelihood evaluations and live-point configurations; see Section \ref{Source3_4Sect} below.
\item {\bf Imperfect sampling constrained-prior:}  this can yield potential systematic error due to incorrect or insufficient exploration of constrained regions. It is also described in  Section \ref{Source3_4Sect} below.
\end{itemize}
%
\subsubsection{Uncertainty in the constrained prior area estimations}\label{Source2Sect}
The constrained prior masses $a_{\max}^{(i)}$ are random variables and we estimate them using their expected value and, with other approximations, we arrive to $\widehat{a}_i= \exp\left(-\frac{i}{N}\right) \approx a_{\max}^{(i)}$.  Since in the standard NS algorithm, the approximation $\exp\left(-\frac{1}{N}\right)$ of $\frac{N}{N+1}$ is used, and since $\exp\left(\mathbb{E}[ \log\widetilde{a}_{\texttt{max}}^{(i)}]\right)=\exp\left(-\frac{1}{N}\right)$, we work here in the logarithmic domain, where we have:
 $$
 \mathbb{E}[ \log\widetilde{a}_{\texttt{max}}^{(i)}]=-\frac{1}{N}, \qquad \operatorname{Var}\left[\log \widetilde{a}_{\texttt{max}}^{(i)}\right]=\frac{1}{N^2}.
 $$
Consequently, following the recursion in Eq. \eqref{RecursionAi}, we also have:
$$
 \mathbb{E}[ \log a_{\texttt{max}}^{(i)}]=-\frac{i}{N}, \qquad \operatorname{Var}\left[\log a_{\texttt{max}}^{(i)}\right]=\frac{i}{N^2},
 $$
starting with $a_{\texttt{max}}^{(0)}=1$. Thus, the uncertainty in the logarithm of the $a_{\texttt{max}}^{(i)}$ is often defined by the formula $\operatorname{std}\left[\log a_{\texttt{max}}^{(i)}\right]=\dfrac{\sqrt{i}}{N}$ \cite{Keeton11,skilling2006nested}.

\subsubsection{Stochastic uncertainty under exact sampling and possible bias}\label{Source3_4Sect}

Even if every replacement point is drawn exactly from the suitable constrained prior, the evidence estimate remains random for two reasons:
\begin{itemize}
\item[-] {\bf Prior-mass uncertainty:} the ratios $\widetilde{a}_{\texttt{max}}^{(i)}$ of areas are random variables, as we have seen above.
\item[-] {\bf Monte Carlo variability:} the sequence of live points and of the discarded likelihood values has also a random nature.
\end{itemize}
These sources of uncertainty are intrinsic to the standard ideal NS procedure. Under the ideal assumption of exact and independent sampling from the constrained priors, in the literature, a classical approximation for the uncertainty in the logarithm of the evidence is
\begin{align}
\operatorname{std}\Big[\log \widehat{Z}_{\texttt{ev}}\big] \approx \sqrt{\frac{H}{N}},
\end{align}
where $N$ is the number of live points and we have denoted as
\begin{align}\label{KLeqPostPrior}
H=\int p(\x |Ê\y) \log \frac{p(\x |\y)}{g(\x)} \mathrm{d} \x=\int \post(\x) \log \frac{\post(\x)}{g(\x)} \mathrm{d} \x = \mathbb{E}_{p(\x | \y)}[\log \ell(\y|\x)]-\log Z_{\texttt{ev}},
\end{align}
is the Kullback-Leibler (KL) divergence (a.k.a., relative entropy) of the posterior with respect to the prior \cite{chopin2010properties,Keeton11}.
The value $H$ measures how much the posterior differs from the prior  (in units of {\it nats} when natural logarithms are used). 
 If the posterior is similar to the prior, $H$ is relatively small. If the posterior is concentrated in a small region of the prior, $H$ is generally larger. Hence, the quantity $e^{-H}$ can be interpreted as an approximate effective fraction of prior mass associated with the posterior concentration.  In other words, it is the amount of information lost when $g(\x)$ is used to approximate  $\post(\x)$. Hence, the factors
$
\operatorname{Var}\Big[\log \widehat{Z}_{\texttt{ev}}\big]  \propto \frac{1}{N}$, and $\operatorname{std}\Big[\log \widehat{Z}_{\texttt{ev}}\big]  \propto N^{-1 / 2}$ that are the familiar Monte Carlo (MC) rates.
\newline
 In practice, the new live point is often generated using MCMC methods (e.g., slice sampling). A NS procedure based on MCMC replacement may fail to explore the full constrained region if it mixes poorly, remains trapped in one mode, or uses too few transitions. In that case, the new added point may not be distributed according to the required constrained prior.
This can introduce {\it systematic error or bias} in the evidence estimate, in addition to the stochastic uncertainty described above.
Repeated independent runs can help assess stochastic variability, while the adequacy of constrained-prior sampling requires separate and more specific diagnostic checks. Consequently, uncertainty estimates based on ideal NS theory should be interpreted cautiously.
}

{
\subsection{NS behavior varying $N$}\label{BehNsect}

Increasing the number of live points $N$ generally improves the statistical accuracy of NS but, clearly, increases the computational cost. Indeed, recall the theoretical result above $\operatorname{std}\left[\log a_{\texttt{max}}^{(i)}\right]=\frac{\sqrt{i}}{N}$: for a fixed iteration $i$, the uncertainty in log prior mass decreases as $1/N$. However, given a suitable stopping rule, larger $N$ also changes the number of iterations required to reach a given prior mass (i.e., given a stopping rule, varying $N$ changes the total number of iterations $I$).   Given the expression $\operatorname{std}\left[\log a_{\texttt{max}}^{(i)}\right]=\frac{\sqrt{i}}{N}$ and a fixed (normalized) area value $a\in(0,1]$, the expected number of iterations needed to reach the value $a$ is approximately:
$$
i^* \approx -  N \lceil \log a\rceil,
$$
where $\lceil a\rceil$ represents  the smallest integer that is greater than or equal to $a$.
The standard deviation of $\log  a_{\texttt{max}}^{(i)}$ at this iteration $i^*$ is then approximately
$$
\operatorname{std}\left[\log a_{\texttt{max}}^{(i^*)}\right] \approx \frac{ \sqrt{-N \lceil \log a\rceil}}{N} = \sqrt{\frac{ -\lceil \log a \rceil}{N}}  .
$$
Thus, at a comparable compression level, the  uncertainty in terms of std decreases again at the familiar Monte Carlo rate $N^{-1 / 2}$, rather than $N^{-1}$ \cite{chopin2010properties,Keeton11}.
As remarked previously, even in the idealized NS procedure,
he computational cost also increases as $N$ grows.
Indeed, even under exact/ideal sampling of the constrained priors, the expected log prior mass after $i$ iterations is approximately $
\mathbb{E}\left[\log a_{\texttt{max}}^{(i)}\right]=-\frac{i}{N}$.
 The posterior density typically assigns substantial mass to regions whose log-prior mass is of the order of $-H$, where $H$ is the KL divergence given in Eq. \ref{KLeqPostPrior} \cite{chopin2010properties,higson2019dynamic,Keeton11}. This motivates the rough estimate 
$$
 \widehat{I}\approx   N \lceil H\rceil,
$$
for the number of iterations needed to reach the region contributing most of the evidence. Moreover, when ideal sampling from the constrained prior distributions is not feasible, the computational cost can increase further, particularly when methods such as rejection sampling are employed. On the other hand, increasing the number of live points $N$ improves the initial coverage of the parameter space and, consequently, the probability of capturing distinct posterior modes. This can be particularly important for multimodal posteriors. Indeed, a mode with very small prior mass may still contribute substantially to the evidence, and hence have a large posterior mass, if the likelihood within that mode is sufficiently high.
\newline
Therefore, increasing $N$ generally improves statistical accuracy, but also increases the number of iterations required to achieve a pre-established level of prior-mass compression.\footnote{{Prior-mass compression refers to the progressive reduction of the remaining prior mass as the likelihood threshold increases, hence restricting the exploration to increasingly smaller regions of the parameter space with higher likelihood values.}} Moreover, a larger number of live points improves the initial coverage of low-prior-mass modes, although it does not guarantee their subsequent exploration. These theoretical considerations must be distinguished from implementation-specific errors due to imperfect constrained-prior sampling \cite{higson2019dynamic,Keeton11}.
}

\section{Relationship with importance sampling (IS) }\label{SuperSect}
{
\subsection{NS and IS estimators of the evidence as weighted likelihood values }
} 
We can rewrite Eq. \eqref{EqNestedZ} as
\begin{equation}\label{EqNestedZ_2}
\boxed{
\begin{aligned}
\widehat{Z}_{\texttt{ev}} &=   \sum_{i=1}^{I}  \underbrace{(a_{\texttt{max}}^{(i-1)} -a_{\texttt{max}}^{(i)} )}_{\gamma_i} \underbrace{\lambda_{\texttt{min}}^{(i)}}_{\ell(\y|\x_{\texttt{rem}}^{(i)})},   \\
\widehat{Z}_{\texttt{ev}} &=   \sum_{i=1}^{I}  \gamma_i \ell(\y|\x_{\texttt{rem}}^{(i)}),	 
\end{aligned}
}
\end{equation}
where we have set $\gamma_i=a_{\texttt{max}}^{(i-1)} -a_{\texttt{max}}^{(i)}>0$ and $\gamma_i \in(0,1)$ for all $i$. Indeed, recall that $a_{\texttt{max}}^{(i)}$ are positive and decreasing values  and $a_{\texttt{max}}^{(i)}\in[0,1]$ (since they represent normalized areas). 
Furthermore, their sum is approximately $1$, i.e.,
\begin{align}
 \sum_{i=1}^{I}  \gamma_i&=(a_{\texttt{max}}^{(0)} -a_{\texttt{max}}^{(1)} )+(a_{\texttt{max}}^{(1)}-a_{\texttt{max}}^{(2)})+(a_{\texttt{max}}^{(2)}-a_{\texttt{max}}^{(3)})+...+(a_{\texttt{max}}^{(I-1)} -a_{\texttt{max}}^{(I)}), \nonumber \\ 
 &=a_{\texttt{max}}^{(0)} - a_{\texttt{max}}^{(I)} , \nonumber \\ 
 &=1- a_{\texttt{max}}^{(I)}.
\end{align}
if $I \rightarrow \infty$, we have  $\sum_{i=1}^{I}  \gamma_i\approx 1$.
where we have used $ \lim\limits_{I \rightarrow \infty} a_{\texttt{max}}^{(I)}=a_{\texttt{max}}^{(\infty)}=0$.
From the expression above, it is apparent that $\widehat{Z}$  is a linear (asymptotically  convex, for $I \rightarrow \infty$) combination of  weights $\gamma_i$.  In \cite{llorenteREV1}, the authors describe an  importance sampling (IS) estimator with a similar form for the marginal likelihood $Z_{\texttt{ev}}$:
\begin{align}\label{aquiZdos}
\fbox{$\displaystyle \widehat{Z}_{\texttt{ev}} = \sum_{i=1}^I \rho_i \ell(\y|\x_i), \qquad \{\x_i \}_{i=1}^I \sim q(\x)$,}
\end{align}
where $q(\x)$ is a generic proposal density, $\rho_i=\dfrac{g(\x_i)}{I q(\x_i)}$. 

{
{\rem   We have already pointed out that  NS is an automatic quadrature scheme since $a_{\texttt{max}}^{(i)}$, $\lambda_{\texttt{min}}^{(i)}$   are  chosen by the NS procedure, as stated in Remark \ref{REM-AutomaticQuad}:  the ordinates $\lambda_{\texttt{min}}^{(i)}$ are  selected by a stochastic procedure, and the values  $a_{\texttt{max}}^{(i)}$ are estimated by the deterministic formula $\widehat{a}_i=\exp\left(-\frac{i}{N}\right)$. Hence, $\gamma_i\approx (e^{-\frac{i-1}{N}}-e^{-\frac{i}{N}})$ are also deterministic weights, as the IS weights that are deterministic $\rho_i=\frac{g(\x_i)}{I q(\x_i)}$, conditioned to $\x_i$.}
}

 {
\subsection{NS estimators of posterior expectations}
} 

{ The NS procedure can be also employed for estimating generic posterior expectations, i.e., $M=\mathbb{E}_{\post}[f(\x)]=\int_{\Theta} f(\x) \bar{\pi}(\x) d\x$ where $\post(\x)=p(\x|\y)=\dfrac{1}{Z_{\texttt{ev}}} \ell(\y|\x)g(\x)$.

{\rem Note that the product NS weights $\gamma_i=\Delta a_{\texttt{max}} =a_{\texttt{max}}^{(i-1)} -a_{\texttt{max}}^{(i)}$ and $\lambda_{\texttt{min}}^{(i)}=\ell(\y|\x_{\texttt{rem}}^{(i)})$, i.e., $\gamma_i \lambda_{\texttt{min}}^{(i)}$, represent a portion of posterior mass: more precisely, a quadrature contribution associated with an interval of prior mass.}}
\newline
Hence, {although we have not a partition of $\x$-space (see Remark \ref{RemNoPartition}),} we can assert that the product $\gamma_i\ell(\y|\x_{\texttt{rem}}^{(i)})$ represents a portion of area below the unnormalized posterior $\pi(\x)$.
 Indeed, the sum of this products provides an approximation of $Z_{\texttt{ev}}=\int_{\Theta}\pi(\x)d\x$, i.e., $\widehat{Z}_{\texttt{ev}}= \sum_{i=1}^I \gamma_i \ell(\y|\x_{\texttt{rem}}^{(i)})$. Thus, the normalized weights 
\begin{align}\label{aquiWpost}
\fbox{$\displaystyle {\bar w}_i= \dfrac{\gamma_i\ell(\y|\x_{\texttt{rem}}^{(i)})}{\widehat{Z}_{\texttt{ev}}}=\dfrac{\gamma_i\ell(\y|\x_{\texttt{rem}}^{(i)})}{\sum_{k=1}^I \gamma_k \ell(\y|\x_{\texttt{rem}}^{(k)})}$,}
\end{align}
allow the approximation of the posterior measure as
\begin{align}\label{postAppro}
\widehat{\pi}(\x)=\sum_{i=1}^I  {\bar w}_i \delta(\x-\x_{\texttt{rem}}^{(i)}).
\end{align}
where $\sum_{i=1} ^I  {\bar w}_i =1$. Hence, as an IS scheme, the NS method can be also applied to approximate generic integral involving the posterior distribution (as a generalized moment), 
$$
M=\mathbb{E}_{\post}[f(\x)]=\int_{\Theta} f(\x) \bar{\pi}(\x) d\x \approx \widehat{M}_{\texttt{NS}}=\frac{1}{\widehat{Z}_{\texttt{ev}}}\sum_{i=1}^I \gamma_i\ell(\y|\x_{\texttt{rem}}^{(i)})f(\x_{\texttt{rem}}^{(i)}) =\sum_{i=1}^I {\bar w}_i f(\x_{\texttt{rem}}^{(i)}),
$$
where is $f(\x)$ is any integrable function w.r.t. $\post(\x)$. { If the constrained-prior sampling is exact,  the estimator above is consistent for $I \rightarrow \infty$. A more refined version for a finite $I$ value, we should include the contribution of the last population of live points as in Eq. \eqref{RefEvidEq}. Following Remark \ref{RefEvidEqREM}, we can write:
\begin{align}
\boxed{
\widehat{M}_{\texttt{NS}}=\frac{\sum_{i=1}^I\gamma_i \ell(\mathbf{y} |\x_{\mathrm{rem}}^{(i)}) f(\x_{\mathrm{rem}}^{(i)})+\frac{ a_{\texttt{max}}^{(I)}}{N} \sum_{n=1}^N \ell(\mathbf{y} |\x_n^{(I)}) f(\x_n^{(I)})}{\sum_{i=1}^I\gamma_i \ell(\mathbf{y} |\x_{\mathrm{rem}}^{(i)})+\frac{ a_{\texttt{max}}^{(I)}}{N} \sum_{n=1}^N \ell(\mathbf{y} |\x_n^{(I)})} .
}
 \label{Ref_IS_gen_Eq}
\end{align}
The numerator estimates the integral with respect to $f$ with respect to the unnormalized posterior $\pi(\x)$, while the denominator estimates the evidence. Including the final live-point terms in both ensures that the same finite-run approximation is used for the numerator and denominator.
 }

 {
\subsection{Equivalent proposal density:  importance nested sampling \cite{cameron2014recursive,feroz2008multimodal}}

Let us compare Eqs. \eqref{EqNestedZ_2} - \eqref{aquiZdos}.}
The NS estimator can be interpreted as a IS estimator of the form \eqref{aquiZdos}, using a ``sophisticated'' proposal density such that 
$\rho_i=\gamma_i=a_{\texttt{max}}^{(i-1)} -a_{\texttt{max}}^{(i)}$. Generally, the  analytical form of this proposal is not available and we cannot evaluate it. However, we can draw from it using the NS procedure, and the weights are computed by the formula $\gamma_i=a_{\texttt{max}}^{(i-1)} -a_{\texttt{max}}^{(i)}$. 
\newline
{
The authors in \cite{feroz2008multimodal} one or more ellipsoidal distributions are used to construct an approximation of the likelihood contours defined by the inequalities $\{\ell(\y|\x)>\lambda\}$. Then, in \cite{cameron2014recursive,feroz2008multimodal}, the authors introduce a {\it pseudo-proposal density} $q^*(\x)$ that is an equivalent proposal density of all generated points in a NS procedure, considering the  ellipsoidal approximations  of the likelihood contours. Hence, the the evidence integral could be approximated using the analytic form of an IS estimator using $q^*(\x)$ as proposal.  This equivalent proposal $q^*(\x)$ can be written as a mixture where each component involves the evaluation of an indicator function, that is $1$  if $\x$  lies within the  corresponding ellipsoidal approximation or $0$ otherwise \cite{cameron2014recursive,feroz2008multimodal}. A brief and good introduction to this topic is also provided in \cite{Williams:2023ppp}.

\subsection{NS-driven importance sampling}\label{NSISsect}
Similarly to the idea introduced in \cite{feroz2008multimodal} with the importance nested sampling,
a possible IS scheme inspired by the NS approach is the following.
Let us consider $N \cdotÊI$ samples,
$$
\mathcal{P}_i=\{\x_n^{(i)}\}_{n=1}^N, \qquad \x_n^{(i)} \sim  q_i(\x|\lambda_i)=\beta g(\x|\lambda_i)+ (1-\beta) q_0(\x), \qquad \beta\in (0,1],
$$
where $g(\x|\lambda_i)$ are the constrained priors, and $q_0(\x)$ is a generic proposal density $\x \in \Theta$. Thus, we can assign the following weights to the samples using the deterministic approach \cite{ElviraMIS15},
\begin{align}
w_n^{(i)}=\frac{\pi(\x_n^{(i)})}{\dfrac{1}{I}\sum\limits_{i=1}^I  q_i(\x_n^{(i)}|\lambda_i)}.
\end{align}
Normalizing them $\bar{w}_n^{(i)}=\frac{w_n^{(i)}}{\sum_{j=1}^I \sum_{k=1}^N w_k^{(j)}}$, we can employ these normalized weights for creating  IS estimators of any posterior expectations \cite{ElviraMIS15,llorente2026optimality}. The values of $\lambda_i$ can be selected according to the NS procedure. We therefore refer to this scheme as \textit{NS-driven importance sampling (NS-IS)}, an adaptive IS method in which the adaptation mechanism is driven by NS. To the best of our knowledge, this approach has not previously been proposed in the literature.
Moreover, a recent work has reformulated nested sampling within the sequential Monte Carlo framework, establishing consistency results for posterior and evidence estimation with MCMC-generated samples, and unbiasedness of the evidence estimator under suitable assumptions \cite{Salomone2025}.

}


{ 
\subsection{The vertical-likelihood perspective given in \cite{polson2014vertical}}

The NS approach shares some similarities with other Monte Carlo methods that exploit {\it vertical}-ordinate information in their sampling procedures, such as {\it slice sampling} sampling and the {\it Swendsen-Wang algorithm} \cite{Liang10,NealSlice}. An interesting common description is given \cite{polson2014vertical}. Let us consider the following latent-variable representation of the likelihood $\ell(\y|\x)$, used in several well-known methods in literature:
\begin{align}
\ell(\y|\x)=\int_0^{\infty} \mathbb{I}\{0<\lambda<\ell(\y|\x)\} d\lambda=\int_0^{\ell(\y|\x)} d\lambda.
\end{align}
where $\mathbb{I}$ is the indicator function defined in Section \ref{Sect_IntegralsRep}. Similar representations of the target density are employed in the fundamental theorem of simulation and the inverse-of-density method \cite[Chapter 2]{martino2018independent}.
We can define an extended posterior $\post(\x,\lambda)$ on the augmented space $\Theta\times\mathbb{R}^{+}$, such that the original posterior,
$\post(\x)=g(\x) \ell(\y|\x) / Z_{\texttt{ev}}$,
is recovered as marginal density with respect to $\lambda$. Specifically, we consider the following joint distribution:
\begin{align}
\post(\x, \lambda)=\frac{\mathbb{I}\{0<\lambda<\ell(\y|\x)\} g(\x)}{Z_{\texttt{ev}}},
\end{align}
where one marginal distribution is exactly $\post(\x)$ and the other marginal is
\begin{align}
 \post(\lambda) 
& =\int_{\Theta} \pi(\x, \lambda) d\x =\frac{1}{Z_{\texttt{ev}}} \int_{\Theta} \mathbb{I}\{\ell(\y|\x)>\lambda\} g(\x) d\x =\dfrac{Z(\lambda)}{Z_{\texttt{ev}}}, \quad \lambda>0,
\end{align} 
where $Z(\lambda)=\int_{\ell(\y|\x)>\lambda} g(\x) d \x$ is the same survival function defined in Eq. \eqref{aquiHlam2}.

{\rem This is the vertical-likelihood representation connected to nested sampling: the marginal density of the auxiliary variable, $\post(\lambda)$, is proportional to the prior mass above the likelihood level $\lambda$.}
\newline
The conditional densities are:
\begin{align}
p(\lambda | \x)=\frac{\mathbb{I}\{0<\lambda<\ell(\y|\x)\}}{\ell(\y|\x)}, \qquad p(\x |\lambda)=\frac{\mathbb{I}\{\ell(\y|\x)>\lambda\} g(\x)}{Z(\lambda)}.
\end{align}
{\rem Note that $p(\x |\lambda)= g(\x |\lambda)$ is exactly the constrained prior distribution, restricted in $\{\x \in \Theta: \ell(\y|\x)>\lambda\}$, used in nested sampling. Moreover, note that $p(\lambda | \x)$ represents the density of uniform distribution $\mathcal{U}([0,\ell(\y|\x)])$. }
\newline
Let now define a weight function  $w: \mathbb{R}^{+} \rightarrow \mathbb{R}^{+}$, and let $W(\lambda)=\int_0^\lambda w(s) d s$ be the corresponding cumulative function. We can also define a modified, weighted extended posterior as:
\begin{align}
\post_w(\x, \lambda)=\frac{w(\lambda) \mathbb{I}\{0<\lambda<\ell(\y|\x)\} g(\x)}{Z_w},
\end{align}
where
\begin{align}
Z_w  &=\int_{\Theta} \int_0^{\infty}  w(\lambda) \mathbb{I}\{0<\lambda<\ell(\y|\x)\} g(\x) d\x d\lambda, \nonumber \\
&=\int_{\Theta}\underbrace{\left[\int_0^{\infty} w(\lambda) \mathbb{I}\{0<\lambda<\ell(\y|\x)\} d\lambda\right]}_{\Omega_w(\x)}  g(\x) d \x=\int_{\Theta} \Omega_w(\x) g(\x) d \x,
\end{align}
and
\begin{align} 
\Omega_w(\x)  &=\int_0^{\infty} w(\lambda) \mathbb{I}\{0<\lambda<\ell(\y|\x)\} d\lambda, \nonumber \\
&=\int_0^{\ell(\y|\x)} w(\lambda) d\lambda=W\big(\ell(\y|\x)\big)= W\circ \ell(\y|\x),
\end{align}
hence $Z_w=\int_{\Theta} W(\ell(\y|\x)) g(\x) d \x$. Moreover, with similar  derivations we can find the two marginal distributions from the joint $\post_w(\x, \lambda)$,
\begin{align}
\post_w(\x)  =\dfrac{g(\x) \ W\big(\ell(\y|\x)\big)}{Z_w}, \qquad \post_w(\lambda)  =\frac{w(\lambda) Z(\lambda)}{Z_w},
\end{align}
and the conditional pdf are in this case:
$$
\begin{aligned} 
p_w(\lambda | \x)=\frac{w(\lambda) \mathbb{I}\{0 \leq \lambda \leq \ell(\y|\x)\}}{W(\ell(\y|\x))}, \qquad p_w(\x | \lambda)=\frac{g(\x) \mathbb{I}\{\ell(\y|\x) > \lambda\}}{Z(\lambda)}, \\ 
 \end{aligned}
$$
so that $p_w(\x | \lambda)=p(\x | \lambda)=g(\x | \lambda)$ is again the constrained prior distribution, used in nested sampling.
\newline
{\bf Vertical importance sampling \cite{polson2014vertical}.} Let assume the ideal case that we are able to draw samples $(\x_n,\lambda_n)$ from $\post_w(\x, \lambda)$. In this scenario, each component is distributed as the corresponding marginals, i.e., $\x_n \sim \post_w(\x)$ and $\lambda_n \sim \post_w(\lambda)$. The idea in \cite{polson2014vertical} is to use the marginal pdf $\post_w(\x)$ as proposal density $q(\x)$ for estimating $Z_{\texttt{ev}}$ via importance sampling. Thus, in this case, 
\begin{align*}
\widehat{Z}_{\texttt{ev}} = \sum_{i=1}^I \rho_i \ell(\y|\x_i), \qquad \{\x_i \}_{i=1}^I \sim \post_w(\x),
\end{align*}
where
\begin{align}\label{RHO_weights}
\fbox{$\displaystyle \rho_i= \dfrac{g(\x_i)}{I \post_w(\x_i)}=\frac{Z_w}{I \ W\big(\ell(\y|\x_i)\big)}, \qquad \{\x_i \}_{i=1}^I \sim \post_w(\x)$,}
\end{align}
where $Z_w$ is also unknown. However, defining the normalized weights: 
\begin{align}\label{RHO_weights_2}
\fbox{$\displaystyle \bar\rho_i= \dfrac{\rho_i}{\sum_{n=1} \rho_n }=
\frac{\frac{1}{W\big(\ell(\y|\x_i)\big)}}{\sum\limits_{n=1}^N \frac{1}{W\big(\ell(\y|\x_n)\big)} }, \qquad \{\x_i \}_{i=1}^I \sim \post_w(\x)$,}
\end{align}
we can then use the final  estimator  $\widehat{Z}_{\texttt{ev}} = \sum_{i=1}^I \bar{\rho}_i \ell(\y|\x_i)$ \cite{llorenteREV1}. We have solved the problem of the unknown value of  $Z_w$, but recall that drawing from $\post_w(\x)$ is not straightforward. This task would generally require an additional Monte Carlo method, such as an MCMC algorithm. 
 
{\rem  Removing the constant values, the weights  $\rho_i \propto 1/W\big(\ell(\y|\x_i)\big)$ are proportional to the inverse of the value $W\big(\ell(\y|\x_i)\big)$. Hence with the 
 approach in \cite{polson2014vertical}, the choice of proposal $q(\x)$ over the original space $\Theta$ is thereby reduced to the one-dimensional weight function $w(\lambda)$, that generated the  function $W(\lambda)=\int_0^\lambda w(s) d s$. }
 
 {\rem This cumulative function $W(\lambda)$ can be employed to deform/change the likelihood values $W\big(\ell(\y|\x)\big)$ and change then properties of the final IS weights,   $\rho_i \propto 1/W\big(\ell(\y|\x_i)\big)$.  }
\newline
\newline
{\bf Special case $w(\lambda)=1$.} With $w(\lambda)=1$, we have:
\begin{itemize}
\item $\post_w(\x, \lambda)=\post(\x, \lambda)$,
\item  $\post_w(\x) =\post(\x)$  (since $W(\lambda)=\lambda$ in this case),
\item $Z_w=Z_{\texttt{ev}}$ and $\post_w(\lambda) =\dfrac{ Z(\lambda)}{Z_{\texttt{ev}}}$,
\item $\rho_i \propto 1/\ell(\y|\x_i)$, with $\x_i\sim \post_w(\x) =\post(\x)$ and then the method reduces to the so-called harmonic-mean estimator \cite{newton1994approximate,raftery2006estimating}.
\end{itemize} 
The harmonic mean estimator of $Z$ can have infinite variance \cite{raftery2006estimating}. This issue arises because samples $\x_i\sim\post(\x)$ can have arbitrarily small likelihood values, $\ell(\y|\x_i)\approx 0$, leading to arbitrarily large values of the corresponding reciprocal-likelihood terms, $\rho_i=1/\ell(\y|\x_i)$. Clearly, reducing the variability of the likelihood values can mitigate this effect. In light of this consideration, it is natural to seek a decreasing function $w(\lambda)$, obtaining a tempering effect as shown in the second special case below.
\newline
{\bf Special case $w(\lambda)=\alpha \lambda^{\alpha-1}$ with $\alpha \in[0,1]$.} In this scenario, $W(\lambda)=\lambda^a$, and the marginal density, used as proposal, is a tempered version of the  posterior:
\begin{align}
\post_w(\x)=\frac{g(\x) \ell(\y|\x)^\alpha}{Z_w}, \qquad \alpha \in[0,1],
\end{align}
hence the method becomes the method described in  \cite[Section 4.3.2]{llorenteREV1}. With $\alpha <1$, the idea is to construct a proposal distribution by manipulating the posterior density so that it will retain a similar shape, but with higher variance or heavier tails. This approach is employed in several well-known techniques, such as the stepping-stones algorithm,  the power-posterior method and path sampling \cite{friel2008marginal,gelman1998simulating},\cite[Section 4]{llorenteREV1}.

{\rem Similarly to the NS-IS scheme described in Section \ref{NSISsect}, and in light of the considerations above, the standard NS procedure can be interpreted as using the constrained prior densities as adaptive proposals for generating the live points, with $\lambda$ acting as the adaptation parameter.   }

}

\section{Practical and theoretical limitations in standard NS} \label{NestedApprox}

In the following, we describe the more critical points in nested sampling:
\begin{itemize}
\item {{\bf Constrained-prior sampling bias.}} Arguably, the most critical task in the implementation of nested sampling is drawing samples from the truncated priors $g(\x|\lambda)$. {Note that the theoretical results rely on the assumption that the live points are independent and identically distributed (i.i.d.) samples from the corresponding constrained priors. If this assumption does not hold, the theoretical foundations of NS may no longer be valid, and the resulting constructions and guarantees may fail.} For drawing from $g(\x|\lambda)$, one can use a rejection sampling or an MCMC scheme. In the first case, we sample from the prior and then accept only the samples $\x'$ such that $\ell(\y|\x')>\lambda$. However, as $\lambda$ grows, its performance deteriorates  since the acceptance probability gets smaller and smaller. The MCMC algorithms could also have poor performance due to the sample correlation, especially when the support of the constrained prior is formed by disjoint regions or distant modes, {as remarked below \cite{chopin2010properties}}. More generally, there are several possible issues: (a) the constrained region can be extremely thin;  (b) in high dimension, most of the prior mass is near the boundary;  and, as previously noticed, (c) the region may be disconnected (multimodal posteriors). If constrained sampling is imperfect, we can add a bias to the estimator. In practice, the performance depends heavily on the internal sampling method. See also Section \ref{SolSect} below.

\item { {\bf Several likelihood modes.}  The problem discussed above becomes even more severe when the likelihood exhibits multiple modes. In this case, the support of the constrained prior $g(\x|\lambda)$ may consist of disconnected regions, as illustrated in Figure \ref{figNesteda}, making it more difficult to sample from $g(\x|\lambda)$. In particular, MCMC algorithms may become trapped within a single region and fail to explore the remaining modes, leading to mode loss.}

\item {{\bf Stochastic uncertainty in the quadrature variables.}} Moreover, in the derivation of the  NS method we have considered different approximations:
\begin{itemize}
\item  
 The value $\widetilde{a}_{\texttt{max}}^{(i)}$ cannot be computed but estimated with the expected value of the Beta distribution $\mathcal{B}(N,1)$, i.e.,  $\mathbb{E}[\mathcal{B}(N,1)]=\frac{N}{N+1}$.
 
\item The  expected value $\mathbb{E}[\mathcal{B}(N,1)]=\frac{N}{N+1}$  is further replaced and approximated with an exponential function $\exp\left(-\frac{1}{N}\right)$ in Eq. \eqref{approxMeanBeta}.  Hence, Eq. \eqref{approxMeanBeta} contains two approximations. This step could be avoided, keeping directly $\frac{N}{N+1}$.
 The simplicity of the final formula $\exp\left(-\frac{i}{N}\right)$ is perhaps the reason of using the approximation $\frac{N}{N+1} \approx \exp\left(-\frac{1}{N}\right)$. 

\item A further approximation $\E[a_{\texttt{max}}^{(i)}]\approx \E[\widetilde{a}_{\texttt{max}}^{(i)}] \E[a_{\texttt{max}}^{(i-1)}]$ is also implicitly applied in \eqref{approxMeanBeta2} (due to the assumption that ``live'' points in the population are independent). Clearly, when MCMC is used for the constrained sampling, this assumption is violated.  Since MCMC-based constrained sampling is the dominant practical approach, study the impact of this approximation would deserve more attention.

\item { Furthermore, the likelihood thresholds $\lambda_{\texttt{min}}^{(i)}$ are determined by a stochastic process that also depends on the number of live points $N$. As discussed in Section \ref{BehNsect}, the choice of $N$ determines a trade-off between grid resolution and computational cost. }
\item 
{ Last but not least,  if the likelihood has a plateau (constant region) at a value $c$, several live points may have exactly the same likelihood $c$. Consequently, the minimum can be attained by multiple points, and the algorithm must decide how to handle the tied points. The likelihood threshold may remain unchanged over several iterations since several points must be removed successively (according to some criterion to handle ties), but still $\lambda_{\min }^{(i)}=c$ until the last point $\x'$ with $\ell(\y|\x')=c$ is removed  (hence, we have the same node  $\lambda_{\min }^{(i)}=c$ for several indices $i$ in the quadrature formula). Moreover,  the standard derivation for finding the areas $\widehat{a}_i$ does not automatically apply in this scenario and requires additional care \cite{Fowlie21plateaus}. }

\end{itemize}
\item {{\bf Possible weight degeneracy.}} NS is primarily an evidence estimator. Posterior weights can be often highly skewed and, as a consequence, the effective sample size (ESS) can be small (early points have often very small weights) \cite{elvira2022rethinking,MartinoESS2017,MartinoESS2025}. So NS posterior sampling may be less efficient than well-tuned Monte Carlo for parameter estimation (such as MCMC or IS schemes).
\end{itemize}
Hence, as a summary, the  values of the areas $a_{\texttt{max}}^{(i)}$ and the corresponding values $\lambda_{\texttt{min}}^{(i)}$ in the NS estimator \eqref{EqNestedZ} are computed approximately (they are analytically intractable/unknown values). Sampling from the likelihood-constrained priors is the crucial point and the main challenge. {
 Recall that other sources of error mentioned in Sections \ref{QuadratureNSsect} and \ref{StocErrSect} are:
\begin{itemize}
\item Deterministic quadrature error (see Section \ref{QuadratureNSsect} and Eq. \eqref{ErrQuadEq}).
\item  Finite-termination and residual-evidence error, as discussed in Section \ref{StopRuleSect2}.
\item Possible implementation-dependent computational error (numerical precision etc.).
\end{itemize}
Finally, if some of the assumptions stated in Section \ref{Sect_IntegralsRep} are not satisfied, the theoretical derivations become {\it even} more challenging, and many of the results must be carefully revisited. For instance, if the likelihood  has constant regions, NS can produce inaccurate estimations of the evidence and posterior densities    \cite{Fowlie21plateaus}.
}

\section{Advanced NS schemes in literature}
\label{SolSect}

Nested Sampling estimators have been extended to a variety of settings. For example, in likelihood-free scenarios - where only an unbiased estimate of the likelihood is available - adapted versions of the NS algorithm have been developed  \cite{Mikelson2020}. The combination with importance sampling has been designed in \cite{chopin2010properties}.
\newline
Furthermore, several strategies have emerged to address the main challenges of NS, such as high-dimensionality, multimodality, and strong parameter degeneracies. In this section, we review the principal variants and approaches. 
 The {\it ellipsoidal NS} technique, as included in MultiNest implementation \cite{feroz2008multinest}, approximates the current set of live points by one or more ellipsoids. New samples are drawn uniformly from the union of these ellipsoids and are accepted if they satisfy the likelihood constraint condition. By employing multiple ellipsoids, the method can effectively handle separated modes. {Other ideas for tackling complex geometry supports and  multimodality are given in \cite{Buchner2021UltraNest}. }
The slice sampling method and the Hamiltonian Monte Carlo (HMC) algorithm has been proposed to be used within NS  \cite{Betancourt_HMC_NS,handley2015polychord}
 The main idea in \cite{Betancourt_HMC_NS} is to leverage HMC dynamics to traverse the constrained likelihood region more efficiently. By utilizing gradient information, it can reduce the random-walk behavior common in traditional MCMC approaches and enables efficient exploration over long distances in parameter space. This method seems to work well in high-dimensional smooth posteriors and is particularly effective for Bayesian models with differentiable likelihoods. Its limitations include the requirement of gradients and the complexity of correctly handling the likelihood boundary defined by the constrain. In addition, Hamiltonian NS is less robust for multimodal distributions. 
\newline
The \textit{diffusive Nested Sampling} (Diff-NS) scheme \cite{brewer2011diffusive} introduces an alternative exploration mechanism in which particles are allowed to diffuse across likelihood levels rather than progressing monotonically toward increasingly constrained regions. Instead of enforcing a strictly inward shrinkage of prior mass, Diff-NS constructs a sequence of likelihood levels that can be explored reversibly. In this framework, particles are not restricted to the current likelihood constraint but may move upward to more constrained levels or downward to less constrained ones, thereby improving mixing across the hierarchy of constrained priors. The up/down move of Diff-NS is achieved stochastically according to Metropolis-Hastings steps. Importantly, the likelihood levels are not defined by all observed likelihood values. Rather, a relatively small number of levels- typically few of them - is maintained. Within each level, the sampled points are used to estimate the average likelihood over the corresponding prior volume, allowing for a more flexible and globally informed approximation of the evidence.
This process is reminiscent of tempering and enhances mixing, especially for strongly multimodal distributions. Diff-NS reduces the risk of missing isolated modes that might be overlooked by traditional NS methods. While it seems particularly effective for posteriors with extreme multimodality, it is more computationally complex and can result in higher variance in evidence estimates compared to conventional approaches. 
\newline
In \cite{Habeck2015}, the author relaxes the strict likelihood constraint by introducing an auxiliary ``demon'' variable that temporarily absorbs surplus likelihood (interpreted as energy), thereby permitting controlled fluctuations around the imposed likelihood threshold. In   \cite{higson2019dynamic}, the authors have proposed the {\it dynamic NS} method, i.e.,  dynamically varying the number of live points throughout the sampling process. This dynamic allocation focuses computational resources on regions where additional samples most effectively reduce uncertainty in the evidence or posterior estimates.  Other recent works combining nested sampling and machine learning can be found in \cite{Williams2021NSNF,Williams:2023ppp}. {Finally, solutions to the presence of  likelihood plateaus are provided in  \cite{Fowlie21plateaus}. Table \ref{tab:NSmethods} summarizes the main advanced NS schemes in the literature. Broadly speaking, note that most advanced NS methods aim to address the challenge of efficiently sampling from the constrained priors, which is  clearly the main computational issue of the standard NS idea. Furthermore, the main review and survey papers on NS, including the present work, are listed in Table \ref{tab:NSreviews}. This comparison also clarifies the scope and contribution of the present work relative to previous reviews.
}

\begin{table}[h!]
{
\begin{center}
\small
\caption{The main advanced NS approaches in the literature.}
\label{tab:NSmethods}
\renewcommand{\arraystretch}{1.5}
\begin{tabular}{p{4.2cm}| p{9.2cm}}
\hline
\textbf{Primary challenge} & \textbf{Key idea} \\
\hline
Exploration of constrained densities
& \textbf{NS with ``Demons''}
\cite{Habeck2015}:
introduce auxiliary demon variables to facilitate sampling
within likelihood-constrained regions. \\
\hline 
Multimodality and mode separation
& See \cite{Betancourt_HMC_NS} using advanced MCMC schemes, and see \textbf{MultiNest} \cite{feroz2008multinest}:
use clustering and ellipsoidal bounds to identify and sample
separate regions of the constrained prior. \\
\hline
Complex geometry and multimodality
& \textbf{UltraNest} \cite{Buchner2021UltraNest}:
use neighbourhood-based regions and adaptive region enlargement
to construct sampling regions that better represent the geometry
of the live points. \\
\hline
High dimensionality
& \textbf{PolyChord} \cite{handley2015polychord}:
use multivariate slice sampling to generate constrained samples
without relying on rejection from bounding volumes. \\
\hline
Adaptive allocation of computational effort
& \textbf{Dynamic NS} \cite{Speagle2020dynesty}:
adaptive number of live points, fixing/given a computational budget. \\
\hline
Exploration across likelihood levels
& \textbf{Diffusive NS} \cite{brewer2011diffusive}:
allow particles to move both up and down a hierarchy of
likelihood-constrained distributions, promoting exploration
and helping particles escape poorly explored regions. \\
\hline
Complex distributions and expensive likelihood evaluations
& \textbf{nessai} \cite{Williams2021NSNF}:
use normalizing flows to learn transformations that facilitate
proposals within the constrained prior. \\
\hline
likelihood with plateaus & several solutions are presented in \cite{Fowlie21plateaus} \\
\hline
\end{tabular}
\end{center}
}
\end{table}

\begin{table*}[h]
\centering
{
\small
\caption{Reviews, surveys, and tutorials on NS.}
\label{tab:NSreviews}
\renewcommand{\arraystretch}{1.5}
\begin{tabular}{p{3.2cm} p{1.2cm} p{3.1cm} p{8.2cm}}
\hline
\textbf{Reference} & {\bf  Year} &
\textbf{Scope} &
\textbf{Main focus and contribution} \\
\hline
\cite{Buchner23} & 2023
&
General methodological review
&
Systematic review of NS algorithms, constrained-prior sampling,
parallelization, termination criteria, uncertainty estimation,
and diagnostic procedures. \\
\hline
\cite{Ashton2022}& 2022
&
General tutorial and review
&
Introduction to NS theory and practical implementation, including
high-dimensional sampling algorithms, applications, best practices,
and limitations. \\
\hline
\cite{Latz2024NSUncertainty}& 2024
&
Uncertainty quantification
&
Studies NS as a probabilistic quadrature method and develops
theoretical results relevant to uncertainty quantification and
rare-event estimation. \\
\hline
\cite{chopin2007,chopin2007b}& 2007 &
Commentary and critical discussion &
Studies different theoretical aspects and practical applicability of NS, highlighting limitations. \\
\hline
\cite{chopin2010properties}& 2010 &
Theoretical and asymptotic analysis &
Studies convergence of the evidence estimator, its error rate, and the effect of parameter dimension; also discusses alternatives to MCMC for generating constrained samples. \\
\hline
This work& 2026
&
Critical guide and theoretical review
&
Examines the theoretical foundations of NS, with particular
emphasis on the complete theoretical derivations, assumptions,
and potential limitations. The connections IS and vertical MC are examined in particular detail.\\
\hline 
\cite{Keskin2023}& 2023 & Application-oriented paper & Illustrates the application of NS to direction-of-arrival estimation,
discussing practical implementation and parameter estimation.  \\
\hline
\cite{Partay2021}& 2021
&
Materials science
&
Reviews adaptations of NS for sampling potential energy surfaces,
calculating partition functions, and estimating thermodynamic
properties, including parallel and variable-cell extensions. \\
\hline
\end{tabular}}
\end{table*}



\section{Conclusions}\label{SectConc}

The Nested sampling (NS) technique has attracted considerable attention and has been widely applied in the literature. {Its widespread application, particularly in cosmology and astronomy, demonstrates the significant impact and practical success of NS. In our view,} NS inherently performs a kind of likelihood optimization, which underlies much of its remarkable success. We consider this feature particularly important, as identifying regions of high likelihood is a critical aspect for the efficiency of all computational sampling methods. Indeed, the use of the gradient on many sampling methods has the same purpose. {Another strength of NS is that its theoretical foundations automatically provide  tools for assessing the algorithm's reliability and diagnosing potential problems, including criteria for determining when to stop the computation. However, the validity of these tools depends on the underlying assumptions being satisfied. }
\newline
The full derivation is complex and relies on several approximations, as discussed in Section~\ref{NestedApprox}. Sampling from the likelihood-constrained priors is a central challenge and is far from straightforward \cite{chopin2007,chopin2007b,chopin2010properties}, with the quality of this constrained sampling being crucial to the method's performance. In this context{, from our perspective,} the widespread success of NS in the literature is indeed remarkable.
In our view, the comprehensive and detailed description presented in this work can not only enhance understanding and appreciation of nested sampling but also serve as a valuable foundation for the development of future improvements and methodological variants. By clarifying both the theoretical principles and the practical challenges of the algorithm, we hope to provide insights that will guide researchers in designing more efficient implementations, exploring novel extensions, and applying NS to an even broader range of scientific problems.
In this sense, this work serves as a tutorial for newcomers to the field and as a critical review for experienced practitioners.
{Moreover, the combinations of nested sampling, importance sampling, and vertical Monte Carlo discussed in this work may provide a foundation for developing novel methods and stimulating further research.}

 {\footnotesize
 \section*{{\footnotesize Acknowledgements}}
 A preprint version of this work can be found at \url{https://arxiv.org/pdf/2606.17916}. No funding was received to support this work.
}




\bibliographystyle{plain}
\bibliography{bibliografia}


\end{document}